\documentclass[prd,aps,nofootinbib,showpacs,floats,letterpaper,floatfix,groupedaddress,eqsecnum]{revtex4}

\usepackage{dcolumn,epsfig}
\usepackage{amssymb,amsmath}

\def\be{\begin{equation}}
\def\ee{\end{equation}}
\def\beq{\begin{eqnarray}}
\def\eeq{\end{eqnarray}}

\def\IL{\relax{\rm I\kern-.18em L}}

\def\f{\frac}

\begin{document}

\title{Mining information from binary black hole mergers: \\ a comparison of
  estimation methods for complex exponentials in noise}
\author{Emanuele Berti} \email{berti@wugrav.wustl.edu}
\affiliation{McDonnell Center for the Space Sciences, Department of Physics, Washington University, St.  Louis,
Missouri 63130, USA}


\author{Vitor Cardoso} \email{vcardoso@phy.olemiss.edu}
\affiliation{Department of Physics and Astronomy, The University of
Mississippi, University, MS 38677-1848, USA \footnote{Also at Centro
de F\'{\i}sica Computacional, Universidade de Coimbra, P-3004-516
Coimbra, Portugal}}

\author{Jos\'e A. Gonz\'alez} \email{jose.gonzalez@uni-jena.de}
\affiliation{Theoretical Physics Institute, University of Jena, Max-Wien-Platz 1, 07743, Jena, Germany}

\author{Ulrich Sperhake} \email{Ulrich.Sperhake@uni-jena.de}
\affiliation{Theoretical Physics Institute, University of Jena, Max-Wien-Platz 1, 07743, Jena, Germany}

\date{\today}

\begin{abstract}
  The ringdown phase following a binary black hole merger is usually assumed
  to be well described by a linear superposition of complex exponentials
  (quasinormal modes). In the strong-field conditions typical of a binary
  black hole merger, non-linear effects may produce mode coupling. Artificial
  mode coupling can also be induced by the black hole's rotation, if the
  radiation field is expanded in terms of spin-weighted spherical harmonics
  (rather than spin-weighted {\em spheroidal} harmonics). Observing deviations
  from the predictions of linear black hole perturbation theory requires
  optimal fitting techniques to extract ringdown parameters from numerical
  waveforms, which are inevitably affected by numerical error. So far,
  non-linear least-squares fitting methods have been used as the standard
  workhorse to extract frequencies from ringdown waveforms. These methods are
  known {\em not} to be optimal for estimating parameters of complex
  exponentials.  Furthermore, different fitting methods have different
  performance in the presence of noise. The main purpose of this paper is to
  introduce the gravitational wave community to modern variations of a linear
  parameter estimation technique first devised in 1795 by Prony: the
  Kumaresan-Tufts and matrix pencil methods.  Using ``test'' damped sinusoidal
  signals in Gaussian white noise we illustrate the advantages of these
  methods, showing that they have variance and bias at least comparable to
  standard non-linear least-squares techniques.  Then we compare the
  performance of different methods on unequal-mass binary black hole merger
  waveforms. The methods we discuss should be useful both theoretically (to
  monitor errors and search for non-linearities in numerical relativity
  simulations) and experimentally (for parameter estimation from ringdown
  signals after a gravitational wave detection).

\end{abstract}


\maketitle


\section{Introduction}

The last year witnessed a remarkable breakthrough in numerical relativity.
Many different groups were finally able to evolve black hole binaries through
the last few cycles of inspiral, merger and ringdown and to extract
gravitational waveforms from the evolutions
\cite{Bruegmann:2003aw,Pretorius:2005gq,Campanelli2006,Baker2006,Herrmann:2006ks,Sperhake:2006cy,Scheel:2006gg,Bruegmann:2006at,Gonzalez:2006md,Szilagyi:2006qy}.
If numerical simulations start out at large enough separation, it should be
possible to match them with Post-Newtonian predictions for the inspiral
signal. This is a very active research area
\cite{Berti:2006bj,Buonanno:2006ui,Baker:2006ha}.  Studies of equal-mass
merger simulations starting from different orbital separations (or produced by
different numerical techniques) show that, independently of the waveform
accuracy in the pre-merger phase, the strong-field waveform has a
``universal'' shape.  Preliminary, ``first order'' explorations of the merger
waveform for equal-mass black hole binaries indicate that the merger phase is
short-lived, lasting only $\sim 0.5-0.75$ gravitational wave (GW) cycles
\cite{Buonanno:2006ui}. After this short merger the signal has the typical
``ringdown'' shape -- i.e., it is well modelled by a superposition of complex
exponentials, known as quasinormal modes (QNMs).

QNMs play a role in any astrophysical process involving stars and black holes.
Oscillations of a relativistic star and of a black hole necessarily produce
GWs \cite{KokkotasSchutz,vish,Leaver:1985ax}. In linear perturbation theory
the metric perturbations can usually be described by a single scalar function,
which, in the ringdown phase, can be written as a superposition of complex
exponentials:
\be\label{rdwave}
\Psi_{lm}(t) = \sum_{n} A_{lmn} 
e^{(\alpha_{lmn}+i \omega_{lmn})t+i \varphi_{lmn}}\,.
\ee 
Here $A_{lmn}$, $\varphi_{lmn}$, $\omega_{lmn}$ and
$\tau_{lmn}=-\alpha_{lmn}^{-1}$ are the mode's amplitude, phase, frequency and
damping time respectively.  The indices $(l\,,m)$ describe the angular
dependence of the signal, and the index $n$ sorts the modes by the magnitude
of their damping time.  The amplitude and phase is determined by the specific
process exciting the oscillations.  Remarkably, in linear perturbation theory
the QNM frequencies $\omega_{lmn}$ and damping times $\tau_{lmn}$ are uniquely
determined by the mass and angular momentum of the black hole: this is a
consequence of the so-called ``no hair'' theorem of general relativity.  For
this reason, accurate QNM measurements could provide the ``smoking gun'' for
black holes and an important test of general relativity in the strong-field
regime (see eg. \cite{bcw}, listing the dominant QNM frequencies for rotating
black holes and providing fits of the numerical results).

In the GW literature it is often claimed that ``the ringdown phase is well
described by (linear) black hole perturbation theory, and that the ringdown
waveform is given by a simple superposition of damped exponentials'' of the
form (\ref{rdwave}).  This statement is based on two tacit assumptions, that
may well be false under the strong-field conditions typical of a binary black
hole merger:

\begin{itemize}
\item[(i)] {\it Non-linearities are negligible}, so that linear perturbation
  theory applies. However, non-linear effects {\it should} be present in
  binary black hole merger waveforms. Model problems show that non-linearities
  are responsible for a systematic shift in the quasinormal frequencies and,
  more interestingly, for mode coupling \cite{Zlochower:2003yh}.

\item[(ii)] {\it Mode coupling is negligible}. Even if assumption (i) above is
  valid and non-linear effects are small, different multipolar components can
  still be coupled. For example, artificial mode coupling can arise because of
  some commonly used approximations in the wave extraction
  process\footnote{Spin-weighted spherical harmonics are often used to
    separate the angular dependence of the Weyl scalars and to read off the
    $(l\,,m)$ multipolar components of the emitted radiation, but the Kerr
    metric is not spherically symmetric. More correctly, one should use
    spin-weighted {\em spheroidal} harmonics \cite{Berti:2005gp}.
    Spin-weighted spherical harmonics can be expressed as linear
    superpositions of spin-weighted spheroidal harmonics, and this leads to
    some artificial mode mixing in the extracted waveforms
    \cite{Dorband:2006gg,Buonanno:2006ui}. By ``artificial'' we mean that the
    effect -- in this case, mode coupling -- is due to approximations we
    introduce in the simulations, rather than being produced by a physical
    agent (such as non-linearities).}.
\end{itemize}

An important open problem in the interpretation of merger waveforms is to
isolate general relativistic non-linearities from artificial effects induced
by unwanted features of the numerical simulations, such as finite-differencing
errors, errors due to the approximate nature of initial data or spurious
rotational QNM coupling induced by the use of spin-weighted spherical
harmonics in GW extraction. All of these effects depend on the details of the
numerical implementation and of the wave extraction procedure.  The ultimate
goal of GW detection is to assess the validity of general relativity in the
strong-field regime, so the relevance of this problem can hardly be
underestimated.  Understanding non-linear phenomena unique to general
relativity is of paramount importance if we want to discriminate Einstein's
theory from alternative theories of gravity, that may give different
predictions in strong-field situations. Since most of the strong-field
waveform can be described as a QNM superposition, strong-field effects could
well show up in the fine structure of this QNM superposition: in particular,
non-linearities may manifest themselves as time variations of the ringdown
frequencies, or as beating phenomena between different QNMs.

In this paper we take a first step towards the solution of this problem.  We
consider intrinsic limitations of the {\it fitting routines} used to extract
quasinormal ringing parameters, exploring the properties of different fitting
techniques for the extraction of ringdown parameters from numerical waveforms.
This issue can be studied independently of the details of any given merger
simulation.  The idea is that, by quantifying the limitations of a given
fitting method, we should be able to isolate {\it systematic uncertainties}
(due to the fact that we are fitting noisy numerical waveforms) from possibly
more interesting physical features of the waveforms themselves.

The problem of estimating the parameters of complex exponentials in noise has
a long history in science and engineering. Damped sinusoidal signals are the
``real-world counterpart'' of the harmonic oscillators used in Fourier
analysis.  Therefore, estimating the parameters of damped sinusoids is of
primary importance in physics.  Examples include (i) speech and audio
modelling \cite{audiom}, (ii) angle-of-arrival estimation of plane waves
impinging on sensor arrays for radar purposes or radio-astronomy, and antennae
array processing \cite{antennae}, (iii) estimation of lifetimes and
intensities in radioactive decay \cite{positronlifetime}, (iv) nuclear
magnetic resonance and computer assisted medical diagnosis \cite{djermoune}.
This list can be extended to include spectroscopy, the study of mechanical
vibrations (including seismic signal processing), the diffusion of chemical
compounds, economics, and so on.

Fits of ringdown waveforms are usually performed using standard non-linear
least-squares methods (see eg.  \cite{Dorband:2006gg,Buonanno:2006ui} and
references therein). These methods are known to fail in estimating parameters
for a {\it sum} of damped exponentials. In this case, minimizing the squared
error over the data requires the solution of highly non-linear expressions in
terms of sums of powers of the damping coefficient. No analytic solution is
available, and these expressions can usually be solved only with a good
initial guess for the parameters \cite{marplebook}. Even for a single damped
exponential, if the initial guess is inaccurate the algorithm often fails to
converge.  Iterative algorithms, such as gradient descent procedures or
Newton's method, have been devised to minimize these non-linear expressions.
Unfortunately the algorithms are computationally very expensive, sometimes
requiring at each step the inversion of matrices of dimension as large as the
number of data samples \cite{iterative}.  Furthermore, gradient descent
algorithms for multimodal equations sometimes fail to converge to the global
minimum.

Computational difficulties with non-linear least-squares methods led to the
development of suboptimal estimation methods based on linear prediction
equations \cite{marplebook,djermoune}.  These methods are modern variations on
a 1795 paper by Gaspard Riche, Baron de Prony \cite{prony}, who introduced a
procedure to {\it exactly} fit $N$ data points by as many purely damped
exponentials as needed.  Modern versions of Prony's method generalize the
original idea to damped sinusoidal models. They also make use of least-squares
analysis to {\it approximately} fit an exponential model for cases where the
data points cannot be fitted by the assumed number of exponential terms.
Unfortunately, these ``modified least-squares Prony methods'' are very
sensitive to numerical noise. Two successful improvements of these methods,
the Kumaresan-Tufts (KT) \cite{kumaresantufts} and matrix pencil (MP)
\cite{matrixpencil} techniques, make use of singular value decomposition to
improve parameter estimation accuracy in the presence of noise. For excellent
reviews of these and other estimation methods we refer the reader to the first
chapter of \cite{djermoune} (in French) and to Marple's book
\cite{marplebook}. A purpose of this paper is to introduce these estimation
methods to the GW community.

The plan of the paper is as follows. To put our problem in perspective, in
Section \ref{waves} we describe the main features of our numerical simulations
of binary black hole mergers and of the resulting waveforms.  In Section
\ref{sec:prony} we summarize the theory behind different estimation methods,
and in Section \ref{routines} we list the numerical algorithms we implemented
in our comparisons. In Section \ref{fit-performance} we compare the
performance of linear estimation methods (MP and KT) against non-linear least
squares methods, by performing Monte Carlo simulations of damped sinusoidal
signals in Gaussian white noise.  In Section \ref{fitmerger}, we apply
different fitting methods to selected waveforms generated by numerical
relativity simulations. Finally we summarize our results and indicate possible
directions for further research.

\section{Numerical simulations}
\label{waves}

In this Section we briefly describe the numerical simulations of non-spinning,
unequal-mass black hole binaries previously presented in
\cite{Gonzalez:2006md}.  The simulations were performed with the {\sc Bam}
code \cite{Bruegmann:2006at} , using the so-called ``moving puncture'' method.
This method, originally introduced in Refs.\,\cite{Campanelli2006,Baker2006},
is now being used almost routinely by various groups to successfully perform
numerical simulations of inspiralling and merging black hole binaries. Our
simulations of the inspiral of unequal-mass black hole binaries are the same
used to study recoil in Ref.\,\cite{Gonzalez:2006md}.  Details of these
simulations, of the numerical setup and of the {\sc Bam} code are given in
Refs.\,\cite{Bruegmann:2006at,Gonzalez:2006md}.  The main purpose of this
paper is to test fitting methods to extract QNMs from the simulations.
Therefore we defer a detailed investigation of the unequal-mass waveforms and
of their physical properties to a forthcoming publication \cite{follow-ups}.

In the absence of spin, a standard conformally flat black hole binary initial
data set is uniquely determined by the parameters $m_1$, $m_2$, $\vec{P}_1$,
$\vec{P}_2$, $D$ which denote, in this order, the bare masses and linear
momenta of the individual holes, and their coordinate separation. The
specification of these parameters allows us to compute the total
Arnowitt-Deser-Misner (ADM) mass $M$ of the system, the orbital angular
momentum $L=(P_1+P_2)D$, the individual black hole masses $M_1$, $M_2$ and the
mass ratio $q=M_1/M_2$. Finally, the sum of the individual holes' spins and of
the orbital angular momentum gives the total ADM angular momentum $J$ of the
system.

In the parameter study under consideration, the mass ratio $q$ was varied
while keeping constant the initial coordinate separation $D=7\,M$ of the black
holes.  The linear momenta of the individual holes then follow from the
requirement that the system represent a quasi-circular configuration. For a
given separation $D$, the momentum of each puncture can thus be calculated to
third-order post-Newtonian accuracy in the Arnowitt-Deser-Misner,
transverse-traceless (ADMTT) gauge. The resulting relation is given by
Eq.\,(64) in \cite{Bruegmann:2006at} and forms the basis for all initial data
sets discussed in this work.  The parameters used in our simulations are
summarized in Table \ref{tab: ini_pars}.
\begin{table}
  \caption{Total ADM mass, angular momentum and initial binding energy
           of the unequal mass binaries. \label{sim-pars}}
  \begin{tabular}{cccc}
  $q$   & $M$ & $J/M^2$ & $-E_{\rm b}/M$ \\
  \hline
  1.00  & 0.9935 & 0.8845 & 0.0140 \\
  1.49  & 1.2422 & 0.8494 & 0.0134 \\
  1.99  & 1.4914 & 0.7870 & 0.0124 \\
  2.48  & 1.7408 & 0.7232 & 0.0114 \\
  2.97  & 1.9904 & 0.6649 & 0.0104 \\
  3.46  & 2.2401 & 0.6132 & 0.0096 \\
  3.95  & 2.4899 & 0.5679 & 0.0089 \\
  \end{tabular}
  \label{tab: ini_pars}
\end{table}

In the notation of Ref.\,\cite{Bruegmann:2006at}, the simulations have been
performed using the $\chi_{\eta=2}$ approach of the moving puncture method
with a number of grid points $i=56,~64,~72$ on the three innermost refinement
levels, respectively.  This corresponds to resolutions of $h=1/45$, $1/51$ and
$1/58$. In this work we use the waveforms resulting from the low-resolution
and high-resolution simulations, using $i=56$ and $i=72$ grid points,
respectively. The wave extraction procedure is based on the Newman-Penrose
formalism.  The Weyl scalar $\Psi_4$ is extracted according to the method
described in Sec.\,III\,A of \cite{Bruegmann:2006at}. In all simulations
presented in this work, $\Psi_4$ is calculated at extraction radius $r_{\rm
  ex}=30M$. Further details of the numerical setup are given in
Ref.\,\cite{Gonzalez:2006md}.

\begin{figure*}[ht]
\begin{center}
\begin{tabular}{cc}
\epsfig{file=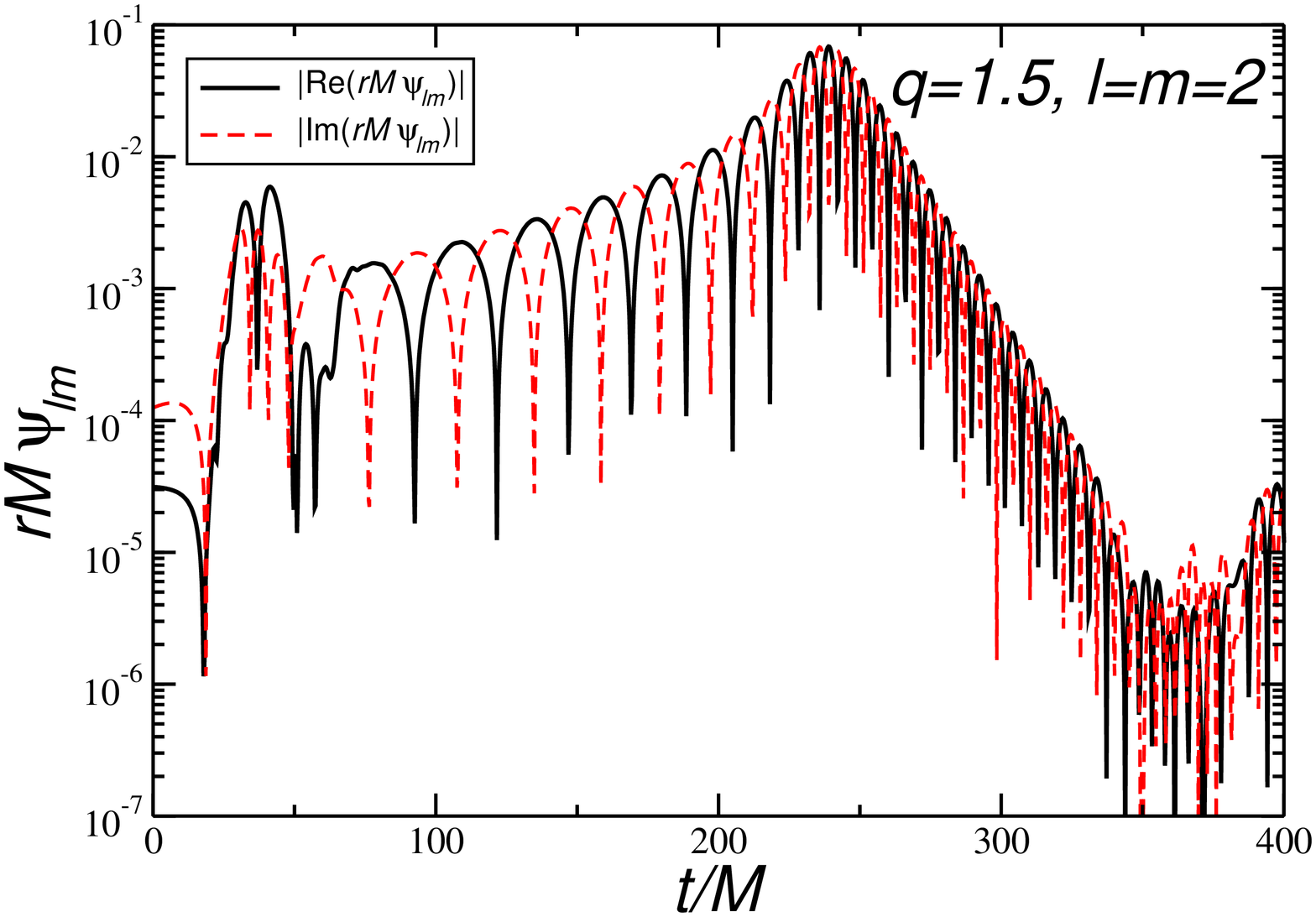,width=7cm,angle=-90} &
\epsfig{file=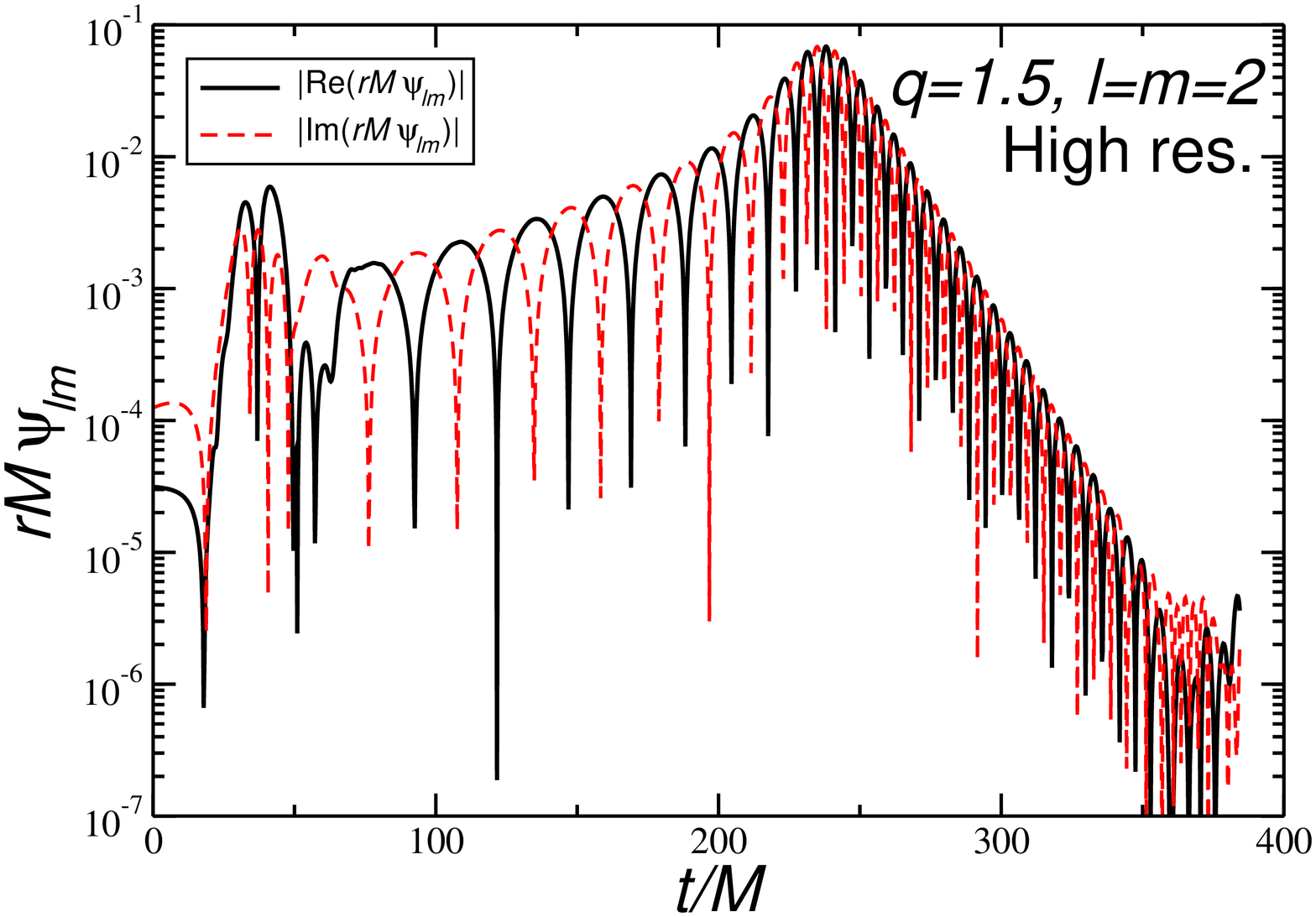,width=7cm,angle=-90} \\
\epsfig{file=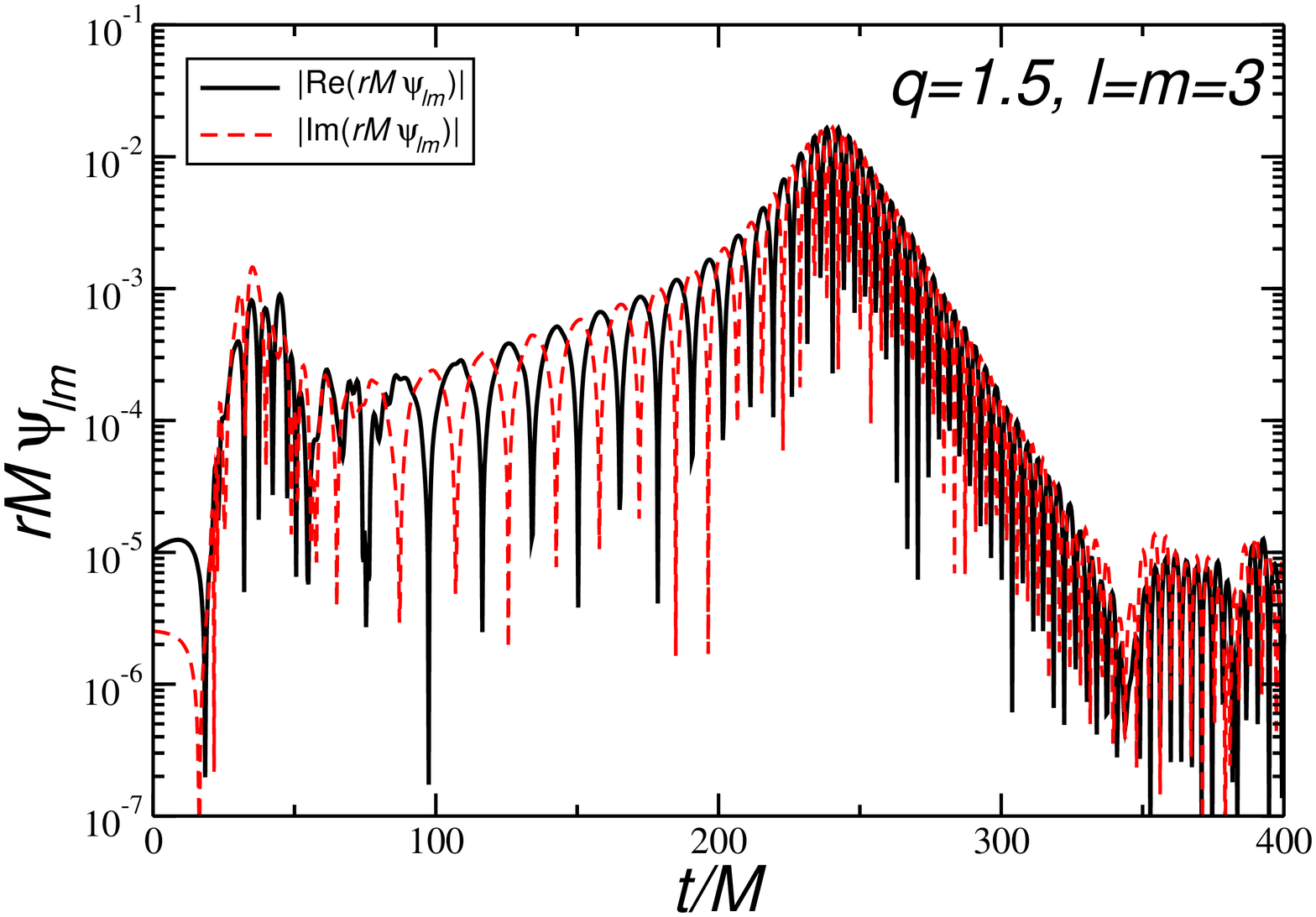,width=7cm,angle=-90} &
\epsfig{file=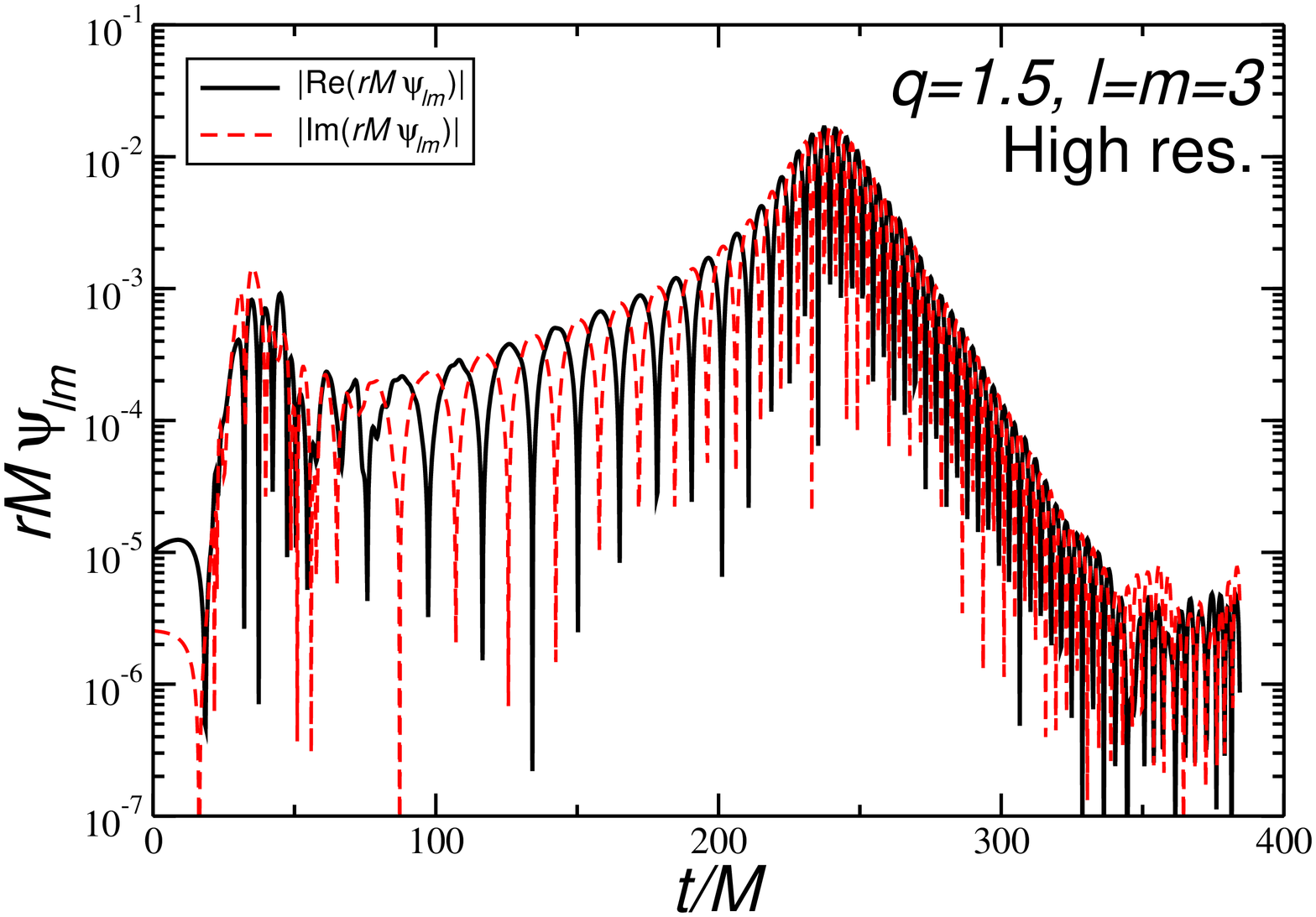,width=7cm,angle=-90} \\
\end{tabular}
\caption{Real and imaginary parts of $rM\psi_{lm}$ for the two dominant
  multipoles. The binary has mass ratio $q=1.5$. The burst of radiation at
  early times is induced by the initial data. After the initial burst, the
  real and imaginary parts are simply related by a phase shift (i.e., the
  waveform is circularly polarized). The irregular behavior at late times is
  due to numerical noise.
  \label{wf1}}
\end{center}
\end{figure*}

\begin{figure*}[ht]
\begin{center}
\begin{tabular}{cc}
\epsfig{file=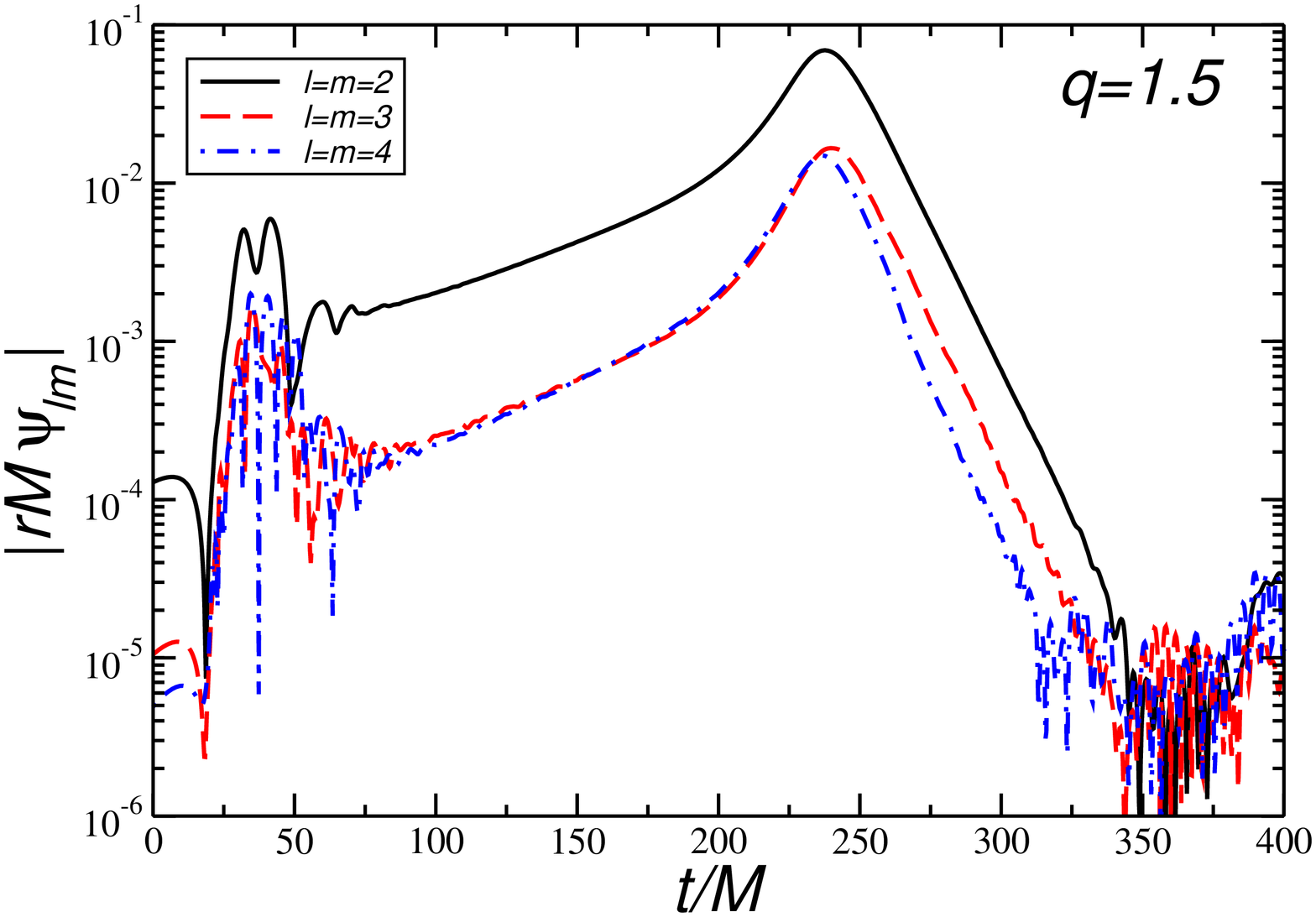,width=7cm,angle=-90} &
\epsfig{file=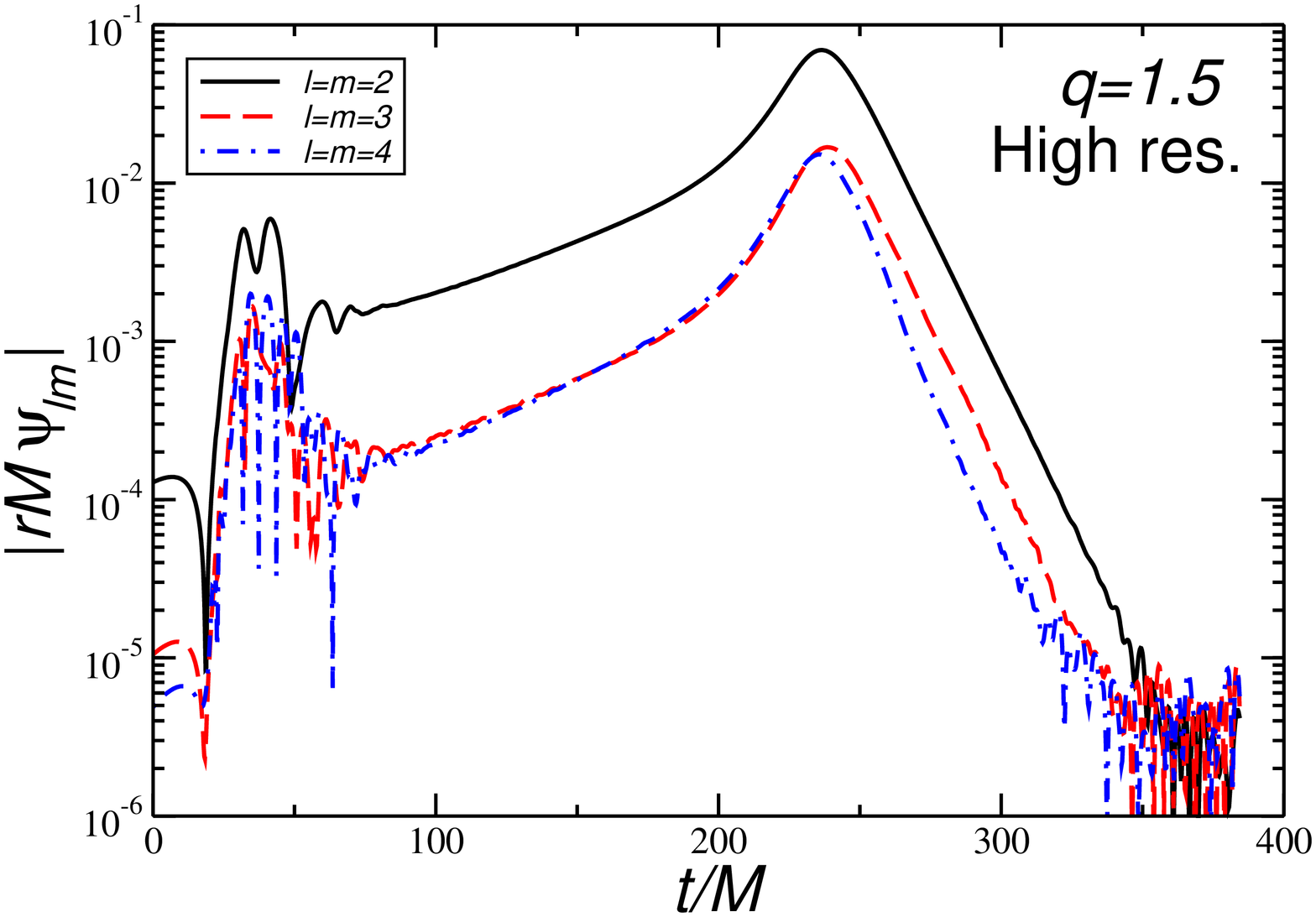,width=7cm,angle=-90} \\
\end{tabular}
\caption{$|rM\psi_{lm}|$ for different multipoles. The high-resolution
  $l=m=2$, $l=m=3$ and $l=m=4$ wave amplitudes have maxima at $t_{\rm
    peak}/M=236.4,~238.4$ and $~235.7$, respectively.  Wiggles at late times
  are due to numerical noise, mainly caused by reflections from the boundaries.
  \label{wf1mod}}
\end{center}
\end{figure*}

To illustrate some issues related with the extraction of ringdown information
from numerical waveforms, in Fig.~\ref{wf1} and Fig.~\ref{wf1mod} we show the
complex mode amplitudes\footnote{These amplitudes are projections of the Weyl
  scalar $\Psi_4$ onto spin-weighted spherical harmonics of spin-weight
  $s=-2$. For the definition see, eg., Eq.~(39) of \cite{Bruegmann:2006at},
  where the amplitudes are denoted by $A_{l m}$.  Here we choose a different
  notation, to avoid confusion with the QNM amplitudes $A_{lmn}$.} $rM
\psi_{lm}$ and their modulus $|rM \psi_{lm}|$ for the dominant components of
the radiation emitted by a binary with mass ratio $q=1.5$\footnote{Here and in
  the following we label the {\sc Bam} runs by the corresponding value of $q$,
  rounded to the first decimal digit (see Table \ref{sim-pars}).}. Because of
the reflection symmetry of the system (see \cite{Bruegmann:2006at}), real and
imaginary parts of the positive-$m$ and negative-$m$ components are related by
\beq
\psi_{l-m}=(-1)^l(\psi_{lm})^*\,.
\eeq 
Components with $|m|<l$ or $l>4$ usually contain significant numerical noise,
and we will ignore them in the following.

At early times the waveform is contaminated by a spurious burst of radiation,
due to the approximate nature of the initial data. After this initial data
burst, the frequency and amplitude of the wave grow as the binary members come
closer, producing the characteristic ``chirping'' gravitational waveform.
Eventually the binary members merge, the wave amplitude reaches a maximum, and
then it decays as the remnant black hole settles down to a stationary Kerr
state, emitting ringdown waves.  The final part of the ringdown waveform is
visibly contaminated by some amount of numerical noise, that decreases as we
increase the resolution. This noise is mainly due to radiation being reflected
from the boundaries of the computational domain.

The plots clearly show that the real and imaginary parts of the waveform
follow the same pattern, except for a (roughly) constant phase shift. This
means that the waveform is circularly polarized \cite{Baker:2002qf}. From the
point of view of extracting information from the ringdown, circular
polarization means that fitting the real or the imaginary part should make no
difference: to a good approximation, we should get the same results for the
oscillation frequencies and damping times. Fitting methods capable of directly
dealing with {\it complex} waveforms (such as the Prony-type methods
considered in this paper) should be particularly useful for waveforms with
general polarization, such as those that should be emitted by the generic
merger of spinning, precessing black holes.

Independently of the chosen fitting method, there is some arbitrariness in the
choice of the time window $[t_0,~t_f]$ used to perform the fit. A well-known
problem with the transition merger-ringdown is that we do not know {\it a
  priori} when the ringdown starts (see
\cite{Buonanno:2006ui,Berti:2006wq,Dorband:2006gg} and references therein).
Ideally, the starting time $t_0$ should be determined by a compromise between
the following requirements: (i) $t_0$ should be small enough to include the
largest possible number of data points: in particular, we do not want to miss
the large amplitude, strong-field part of the waveform right after the merger;
(ii) $t_0$ should be large enough that we do not include parts of the waveform
which are {\em not} well described by a superposition of complex exponentials:
the inclusion of inspiral and merger in the ringdown waveform would produce a
bias in the QNM frequencies.

A judicious choice of $t_f$ is also necessary. Usually we would like the time
window to be as large as possible, but a glance at Fig.~\ref{wf1} and
Fig.~\ref{wf1mod} shows that the low amplitude, late-time signal is usually
dominated by numerical noise, mainly caused by reflections from the boundaries.
This noise can reduce the quality of the fit, especially for the subdominant
components with $l>2$ and for large values of $t_0$.  A practical criterion
for the choice of $t_f$ is suggested by a look at Fig.~\ref{wf1mod}. If the
ringdown waveform were not affected by noise from boundary reflections,
$|rM\psi_{lm}|$ should decay linearly on the logarithmic scale of the
plots\footnote{With larger resolution and longer running times, eventually the
  exponential decay should turn into the well known power-law tail induced by
  backscattering of the radiation off the spacetime curvature
  \cite{Price:1971fb}. In the simulations we consider, noise produced by
  boundary effects is large enough that this effect is not visible.}. At low
signal amplitudes, we see boundary noise-induced wiggles superimposed on this
linear decay: the first occurence of these wiggles is a good indicator of the
time $t_f$ at which numerical results cannot be trusted anymore. To test the
robustness of fitting results to late-time numerical noise, while at the same
time keeping the largest number of data points in the waveform, we decided to
use two different ``cutoff criteria'':

\begin{itemize}

\item[1)] ``Relative'' cutoff: remove from the waveforms all data for times
  $t>t_f=t_{\rm rel}$, where $t_{\rm rel}$ is the time when the amplitude of
  each multipolar component $|rM\psi_{lm}|$ becomes less than some factor
  $\psi_{\rm cutoff}$ times the peak amplitude (at $t_{\rm peak}\sim 240M$ for
  the waveforms in Fig.~\ref{wf1mod}):

  \be 
  \f{|rM\psi_{lm}(t_{\rm rel})|}{|rM\psi_{lm}(t_{\rm peak})|}<
  \psi_{\rm cutoff}\,.  
  \ee

\item[2)] ``Absolute'' cutoff: remove from the fit all data with $t>t_f=t_{\rm
    abs}$, where $t_{\rm abs}$ is the time at which the {\em absolute} value
  of the amplitude $|rM\psi_{lm}|<\psi_{\rm cutoff}/10$.

\end{itemize}

The choice of the cutoff amplitude is somewhat arbitrary. We chose $\psi_{\rm
  cutoff}=10^{-3}$ for low resolution, and $\psi_{\rm cutoff}=10^{-4}$ for
high resolution.

For each chosen $t_f$, we will look at the performance of the fitting routines
as we let $t_0$ vary in the range $[t_{\rm peak},~t_f]$. We do this for two
reasons.  The first reason is physical: by monitoring the convergence of the
QNM frequencies to some ``asymptotic'' value as $t_0\to \infty$, we can tell
if the black hole settles down to a stationary Kerr state, or if, on the
contrary, non-linearities and mode coupling are always present.  The second
reason is related with the main goal of this paper, which is to assess the
performance of our fitting routines: as $t_0$ grows the signal amplitude
decreases exponentially, and we effectively reduce the signal-to-noise ratio
(SNR) in our fitting window.  Robust fitting methods should give reasonable
results even for large values of $t_0$ (that is, modest values of the SNR).

\section{A brief survey of estimation methods for damped sinusoids}
\label{sec:prony}

In this Section we present a brief summary of the theory behind different
estimation methods for damped sinusoidal signals. We consider only methods
that make direct use of the data, by which we mean that no use is being made
of the autocorrelation function. The measured signal $x$ is a linear
superposition of the ``true'' waveform, $\Psi$, and noise, $\varpi$. We
suppose we have $N$ samples of the signal, equally spaced in time with time
sampling interval $T$, and we label each sample by an integer $n$:
\be x[n]= \Psi[n]+\varpi [n]\,,\quad n=1\,,...\,,N\,. \ee
For simplicity we assume the noise $\varpi [n]$ to be white and Gaussian, with
standard deviation $\sigma$ and mean $\mu=0$. We are interested in the
ringdown waveform $\Psi[n]$, that we write as a superposition of $p$ complex
exponentials with arbitrary amplitudes and phases:
\be \Psi[n]=\sum_{k=1}^p A_ke^{(\alpha_k+i \omega_k)(n-1)T+i\varphi_k}\,.
\label{signal}
\ee
It is useful to re-cast Eq.~(\ref{signal}) in the slightly different form
\be \Psi[n]= \sum_{k=1}^p h_kz_k^{n-1}\,,\label{rdwf}\ee
with
\beq h_k&=&A_ke^{i\varphi_k}\,,\label{hprony}\\
z_k&=&e^{(\alpha_k+i \omega_k)T}\,.\label{zprony} \eeq
Now the unknown complex parameters are $\{h_k\,,z_k\}$ ($k=1,\dots,p$), and
possibly the number $p$ of damped sinusoids.

\subsection{Non-linear least-squares}

A popular estimation method is non-linear least-squares, which consists in
minimizing the integrated squared error
\be \rho\equiv \sum_{n=1}^N |\varpi[n]|^2\,,\label{expsqerr} \ee
with $\varpi[n]=x[n]-\Psi[n]$.  The method is very general, in the sense that
(in principle) it can be applied to any model function $\Psi[n]$.  For
ringdown waveforms of the form (\ref{rdwf}) we must minimize over the
$\{h_k\,,z_k\}$ parameter space (and possibly the $p-$space, if the number of
damped exponentials is not known {\it a priori}) in an essentially non-trivial
way.  The procedure is computationally expensive and not always accurate: the
solution may converge to a local (rather than a global) minimum of the
integrated squared error.  Experiments show that non-linear least-squares
techniques perform badly in estimating the parameters of damped exponentials.
The situation gets even worse in the presence of noise
\cite{marplebook,djermoune}. In our numerical work, to minimize the integrated
squared error we used the well-known Levenberg-Marquardt algorithm \cite{LM}
as implemented in the Fortran subroutine \url{lmdif}, which is part of the
\url{MINPACK} library for solving systems of non-linear equations
\cite{MINPACK}\footnote{The same algorithm is also implemented in Mathematica
  \cite{wolfram}.}.

\subsection{Prony method}

Computational difficulties with non-linear least-squares methods led to the
development of suboptimal estimation methods, especially designed to deal with
damped sinusoidal signals. The prototype of these estimation techniques is the
Prony method, which is essentially a trick to reduce the non-linear
minimization problem to a linear prediction problem. The method has been
successfully tested in different branches of data analysis and signal
processing. Some variants of the basic idea improve the variance and bias of
parameter estimation in the presence of noise: we briefly discuss these
variants in the following Sections.  Our introduction to standard Prony
methods parallels those by Marple \cite{marplebook} and Djermoune
\cite{djermoune}. We refer the reader to those references for further details.

To start with, let us assume there is no noise in our data set. Let us also
assume that there are as many data samples as there are complex exponential
parameters $(h_1,\dots,h_p)$, $(z_1,\dots,z_p)$: $N=2p$. We are then fitting
$x[n]$ to an exponential model,
\be x[n]= \sum_{k=1}^p h_kz_k^{n-1} \label{prony1}\,. \ee
For $1\leq n\leq p$ we can write this in matrix form as
\begin{eqnarray}
\left(
  \begin{array}{ccccc}
 z_1^0 & z_2^0 & \cdots & z_p^0\\
 z_1^1 & z_2^1 & \cdots & z_p^1\\
  \vdots & \vdots & \ddots & \vdots\\
  z_1^{p-1} &z_2^{p-1}&\cdots&z_p^{p-1}
  \end{array}
\right) \left(
\begin{array}{c}
h_1 \\
h_2\\
\vdots \\
h_p\\
\end{array}
\right)
= \left(
\begin{array}{c}
x[1] \\
x[2]\\
\vdots \\
x[p]\\
\end{array}
\right)
{}\,.\label{matrixProny1}
\end{eqnarray}
If we can determine the $z_k$'s by some other procedure, then
(\ref{matrixProny1}) is a set of linear equations for the complex amplitudes
$h_k$. Prony's method is in essence a trick to determine the $z_k$'s without
the need for non-linear minimizations, as follows. Let us define a polynomial
${\bf A}(z)$ of degree $p$ which has the $z_k$'s as its roots:
\be {\bf A}(z)=\prod_{k=1}^{p}(z-z_k)\equiv \sum_{m=0}^p a[m]z^{p-m}\,,
\label{pronypolA} \ee
Let us also normalize the complex coefficients $a[m]$ so that $a[0]=1$.  By
using Eq.~(\ref{prony1}) we can write
\be a[m]x[n-m]=a[m]\sum_{k=1}^p h_k z_k^{n-m-1}\,. \ee
Summing this last equation from $m=0$ to $m=p$ we get, for $p+1\leq n \leq 2p$
\be \sum_{m=0}^p a[m]x[n-m]=\sum_{i=1}^p h_iz_i^{n-p}\sum_{m=0}^p
a[m]z_i^{p-m-1}=0\,. \label{zeropol} \ee
where the last equality follows from (\ref{pronypolA}). This is a {\em
  (forward) linear prediction equation}. In matrix form it can be expressed as
\begin{eqnarray}
\left(
  \begin{array}{ccccc}
 x[p] & x[p-1] & \cdots & x[1]\\
 x[p+1] & x[p] & \cdots & x[2]\\
  \vdots & \vdots & \ddots & \vdots                   \\
 x[2p-1] &x[2p-2]&\cdots&x[p]
  \end{array}
\right) \left(
\begin{array}{c}
a[1] \\
a[2]\\
\vdots \\
a[p]\\
\end{array}
\right)
=- \left(
\begin{array}{c}
x[p+1] \\
x[p+2]\\
\vdots \\
x[2p]\\
\end{array}
\right)
{}\,.\label{matrixProny2}
\end{eqnarray}
This equation is the basic result of the Prony method: it shows that we can
determine the $a[k]$'s from the data, decoupling the problem of determining
the $h_k$ and $z_k$ parameters. 

The original Prony procedure works as follows. First, given the data, solve
(\ref{matrixProny2}) for the coefficients $a[m]$. Then determine the roots
$z_k$ of the polynomial (\ref{pronypolA}). The damping and frequency are
obtained by inverting (\ref{zprony}):
\begin{subequations}
\beq \alpha_k&=&\log|z_k|/T\, \,, \\
\omega_k&=&\tan^{-1}\left[{\rm Im}(z_k)/{\rm Re}(z_k) \right ]/T\, \,.\eeq
\end{subequations}
Finally, solve (\ref{matrixProny1}) for the complex quantities $h_k$ and
invert (\ref{hprony}) to find amplitudes and phases:
\begin{subequations}
\beq 
A_k&=&|h_k| \,,\\
\varphi_k&=&\tan^{-1}\left[{\rm Im}(h_k)/{\rm Re}(h_k) \right ]\,.
\eeq
\end{subequations}
It should now be clear that the Prony algorithm reduces the non-linear fitting
problem to the trivial, computationally inexpensive numerical tasks of (i)
solving linear systems of equations, and (ii) finding the roots of a
polynomial.

\subsection{Modified least-squares Prony}

For most situations of interest, there are more data points than there are
exponential parameters: $N>2p$. In this case, Eq.~(\ref{matrixProny2}) is
modified to
%
%
\begin{eqnarray}
\left(
  \begin{array}{ccccc}
 x[p] & x[p-1] & \cdots & x[1]\\
 x[p+1] & x[p] & \cdots & x[2]\\
  \vdots & \vdots & \ddots & \vdots\\
 x[N-1] &x[N-2]&\cdots&x[N-p]
  \end{array}
\right) \left(
\begin{array}{c}
a[1]\\
a[2]\\
\vdots \\
a[p]\\
\end{array}
\right)
= -\left(
\begin{array}{c}
x[p+1] \\
x[p+2]\\
\vdots \\
x[N]\\
\end{array}
\right)
{}\,.\label{matrixProny3}
\end{eqnarray}
We can write this as a matrix equation,
\be {\bf Xa}=-{\bf x} \,,\label{matrixY}\ee
which can be solved in the least-squares sense
\be {\bf a}=-\left( {\bf X}^H {\bf X} \right )^{-1}{\bf X}^H {\bf x}\,, \ee
where a superscript $H$ denotes Hermitian transpose: ${\bf X}^H=({\bf
  X}^T)^*$.  The important difference is that we must now minimize the {\it
  linear prediction} squared error, rather than the {\it exponential
  approximation} squared error of Eq.~(\ref{expsqerr}). Once we have
determined the $a[k]$'s, the rest of the Prony method carries on in the same
manner. Details of the implementation of this ``least-squares Prony method''
(which is sometimes called ``extended Prony method'') can be found in
\cite{marplebook}.

\subsection{Kumaresan-Tufts}

Unfortunately, the original and least-squares Prony methods only work well in
the absence of noise, or for large SNR. For small SNR, the estimation of the
frequencies and damping times has large variance and bias
\cite{marplebook,marcosbook}.

In the previous Section we explained how to estimate the exponential
parameters by introducing the characteristic polynomial ${\bf A}(z)$, which
has roots at $z_k\equiv e^{s_k}=e^{(\alpha_k +i\omega_k)T}$. The coefficients
$a[k]$ of ${\bf A}(z)$ are solutions of the forward linear prediction equation
(\ref{zeropol}). These same exponentials can be generated in reverse time by
the backward linear predictor \cite{marplebook}
\be \sum_{m=0}^p b[m]x[n-p+m]=0\,. \label{zeropol2} \ee
in which $b[0]=1$. The characteristic polynomial
\be {\bf B}(z)=\sum_{m=0}^p b^*[m]z^{p-m}\,,\label{pronypolB} \ee
has roots at $z_k=e^{-s^*_k}$. For damped sinusoids (${\rm Re}[\alpha_k]<0$)
it can be shown that the roots of the forward linear prediction polynomial
${\bf A}(z)$ lie {\it inside} the unit $z-$plane circle, whereas those of the
backward linear prediction polynomial ${\bf B}(z)$ lie {\it outside} the unit
$z-$plane circle \cite{kumaresantufts,marplebook}.

Suppose now we superimpose to the signal complex additive Gaussian white
noise. Noise causes a bias in the estimates of the true zeros of the
polynomials, which translates into a bias in the estimates of frequencies and
damping times. It was empirically observed that the bias may be significantly
reduced by looking for a number of exponential components $L>p$, where $p$ is
the actual number of exponentials present in the signal \cite{kumaresantufts}.
$L$ is called the {\em prediction order} of the model.  Of course, the
selection of a higher prediction order introduces additional zeroes due to
noise, but these can be statistically separated by examining the zeroes of
${\bf A}(z)$ and ${\bf B}(z)$. For both polynomials, zeroes due to the noise
tend to stay within the unit circle, whereas the true zeroes due to the
exponential signal form complex-conjugate pairs inside and outside the unit
circle. This is basically due to the fact that the statistics of a stationary
random process do not change under time reversal. Using singular value
decomposition (SVD) can provide further improvement
\cite{kumaresantufts,marplebook,djermoune}.  Write ${\bf X}$ in
(\ref{matrixY}) as
\be {\bf X}={\bf USV}^H \,,\ee
where ${\bf S}$ is a $(N-L)\times L$ dimensional matrix with the singular
values on the diagonal
$(s_1\,,...\,,s_p\,,s_{p+1}\,,...\,,s_L)$ arranged in
decreasing order. Noise can be reduced by considering the reduced rank
approximation
\be {\bf \hat X}={\bf U{\hat S}V}^H \ee
with
\begin{eqnarray}
{\bf \hat S}=\left[
  \begin{array}{ccccc}
 {\bf S}_p &  {\bf 0}\\
 {\bf 0} & {\bf 0}
  \end{array}
\right]_{(N-L)\times L}
%
\,.\label{matrixreducedrank4}
\end{eqnarray}
Here ${\bf S}_p$ is the top-left $p\times p$ minor of ${\bf S}$. An estimate
for the coefficients $a[k]$ is then
\be {\bf \hat a}=-{\bf \hat X}^{+}{\bf x}\,, \ee
where ${\bf \hat X}^{+}$ is the pseudo-inverse of ${\bf \hat X}$. It was found
\cite{kumaresantufts} that the use of the truncated SVD greatly enhances the
SNR, providing a better estimate of the vector ${\bf \hat a}$, and
consequently of the exponential parameters. This is the basic idea underlying
the Kumaresan-Tufts method. More details can be found in the original work
\cite{kumaresantufts} (see also \cite{marplebook,djermoune}). For improvements
of this method using total least-squares, see for instance \cite{ry}.

The frequency variance depends on the prediction order $L$.  For one {\em
  undamped} exponential ($\alpha=0$) of amplitude $h_1$ it can be shown
\cite{matrixpencil,predictionorder} that the variance is given by
\begin{subequations}
\beq
{\rm var}(\omega)&=&
\frac{2(2L+1)}{3(N-L)^2 L (L+1)}
\frac{\sigma^2}{|h_1|^2}\,
\quad {\rm for} \quad L\leq N/2\,,
\\
{\rm var}(\omega)&=&
\frac{2[-(N-L)^2+3L^2+3L+1]}{3(N-L)L^2 (L+1)^2}
\frac{\sigma^2}{|h_1|^2}\,
\quad {\rm for} \quad L\geq N/2\,.
\eeq
\label{varKT}
\end{subequations}

Minima are attained for $L\simeq N/3$ ($L\leq N/2$) and for $L\simeq 2N/3$
($L\geq N/2$), these relations being exact in the limit $N\to \infty$.
Correspondingly, the optimal frequency variance is
\be {\rm var} (\omega)\simeq
\frac{27}{4N^3}\frac{\sigma^2}{|h_1|^2}\,, \ee
which is only slightly larger than the Cramer-Rao bound:
\be {\rm CRB} (\omega)=\frac{6}{N(N^2-1)}\frac{\sigma^2}{|h_1|^2}\,,
\ee
the ratio being $9/8=1.125$ in the limit $N\to \infty$.  For one damped
exponential, closed-form expressions for the variance and bias of the damping
time in the large SNR limit can be found in
\cite{KTperformance,djermouneKTMP}.

\subsection{Matrix Pencil}

An alternative to estimate exponential parameters from noisy signals is the
matrix pencil (MP) method \cite{matrixpencil}. The MP method is in general
more robust than the KT method, having a lower variance on the estimated
parameters, but a slightly larger bias \cite{djermoune,djermouneKTMP}. For a
detailed description of this method we refer the reader to
\cite{matrixpencil}.  The following brief summary follows quite closely the
treatment in \cite{djermoune}. Let ${\bf X}_0$ and ${\bf X}_1$ be two matrices
defined as
\begin{eqnarray}
{\bf X}_0=\left(
  \begin{array}{ccccc}
 x[0] & x[1] & \cdots & x[L-1]\\
 x[1] & x[2] & \cdots & x[L]\\
  \vdots & \vdots & \ddots & \vdots                   \\
 x[N-L-1] &x[N-L]&\cdots&x[N-2]
  \end{array}
\right)\,;\quad
{\bf X}_1=\left(
  \begin{array}{ccccc}
 x[1] & x[2] & \cdots & x[L]\\
 x[2] & x[3] & \cdots & x[L+1]\\
  \vdots & \vdots & \ddots & \vdots                   \\
 x[N-L] &x[N-L]&\cdots&x[N-1]
  \end{array}
\right) \,.\label{mp1}
\end{eqnarray}
Here $L$ is the pencil parameter: it plays the role of the prediction order
parameter in the KT method. One can decompose ${\bf X}_0$ and ${\bf X}_1$ as
\beq {\bf X}_0={\bf Z}_l {\bf H}{\bf Z}_r\,,\\
 {\bf X}_1={\bf Z}_l {\bf H}{\bf Z}{\bf Z}_r\,,\eeq
where
\begin{eqnarray}
& & {\bf Z}_l=\left(
  \begin{array}{ccccc}
 1 & 1 & \cdots & 1\\
 z_1 & z_2 & \cdots & z_p\\
  \vdots & \vdots & \ddots & \vdots                   \\
 z_1^{N-L-1} &z_2^{N-L-1}&\cdots&z_p^{N-L-1}
  \end{array}
\right)\,;\quad
{\bf Z}_r=\left(
  \begin{array}{ccccc}
 1 & z_1 & \cdots & z_1^{L-1}\\
 1 & z_2 & \cdots & z_2^{L-1}\\
  \vdots & \vdots & \ddots & \vdots                   \\
 1 &z_p&\cdots&z_p^{L-1}
  \end{array}
\right) \,,
\label{mp2}
\end{eqnarray}
\begin{eqnarray}
{\bf H}&=&{\rm diag}(h_1\,,h_2\,,...\,,h_p)\,,\\
{\bf Z}&=&{\rm diag}(z_1\,,z_2\,,...\,,z_p)\,.
\end{eqnarray}
Consider now the matrix pencil\footnote{Let $A$ and $B$ be two $n \times n$
  matrices. The set of all matrices of the form $A-\lambda B$, with $\lambda$
  complex, is called a {\it matrix pencil}.} ${\bf X}_1-z{\bf X}_0$. We can
write this as
\be {\bf X}_1-z{\bf X}_0=
{\bf Z}_l{\bf H}\left ({\bf Z}-z{\bf I}_p\right ){\bf Z}_r\,.
\ee
When $z\neq z_i$, the matrix ${\bf Z}-z{\bf I}_p$ is of rank $p$. However, for
$z=z_i$ it is of rank $p-1$.  Therefore the poles of the signal reduce the
rank of the matrix pencil for $p\leq L\leq N-p$. This is equivalent to saying
that the poles $z_i$ are the generalized eigenvalues of $({\bf X}_1\,,{\bf
  X}_0)$, in the sense that $({\bf X}_1-z{\bf X}_0){\bf v}=0$, with ${\bf v}$
an eigenvector of ${\bf X}_1-z{\bf X}_0$. To find the poles $z_i$ one can use
the fact that ${\bf X}_0^{+}{\bf X}_1$ has $p$ eigenvalues equal to the poles
$z_i$, and $L-p$ null eigenvalues \cite{djermoune}. Here a dagger denotes the
(Moore-Penrose) pseudo-inverse.

In practice we do not have access to the noiseless signal, therefore we must
work directly with the noisy data. SVD is again necessary to select the
singular values due to the signal. The basic steps of the MP method can be
summarized as follows:

\begin{itemize}

\item[(i)] Build the matrices ${\bf Y}_0$ and ${\bf Y}_1$ as in (\ref{mp1});

\item[(ii)] Make a SVD of ${\bf Y}_1$: ${\bf Y}_1={\bf USV}^T$.

\item[(iii)] Estimate the signal subspace of ${\bf Y}_1$ by considering the
  $p$ largest singular values of ${\bf S}$: ${\bf \tilde Y}_1={\bf U}_p{\bf
    S}_p{\bf V}_p^H$, where ${\bf U}_p$ and ${\bf V}_p$ are built from the
  first $p$ columns of ${\bf U}$ and ${\bf V}$, and ${\bf S}_p$ is the top-left
  $p\times p$ minor of ${\bf S}$.

\item[(iv)] The matrix ${\bf Z}_L={\bf Y}_1^+ {\bf Y}_0={\bf V}_p{\bf
    S}_p^{-1}{\bf U}_p^T{\bf Y}_0$ has $p$ eigenvalues which provide estimates
  of the inverse poles $1/z_i$; the other $L-p$ eigenvalues are zero. Since
  ${\bf Z}_L$ has only $p$ non-zero eigenvalues, it is convenient to restrict
  attention to a $p\times p$ matrix ${\bf Z}_p={\bf S}_p^{-1}{\bf U}_p^T{\bf
    Y}_0{\bf V}_p$ \cite{matrixpencil}.

\end{itemize}

The MP technique exploits the matrix pencil structure of the underlying
signal, rather than the prediction equations satisfied by it. Nevertheless,
there are strong similarities between the MP and KT methods. An extensive
comparison of their performance and theoretical properties can be found in
\cite{matrixpencil,djermouneKTMP,djermoune}.

As for the KT method, the variance of parameter estimation depends on $L$.
For one {\em undamped} exponential of amplitude $h_1$ it can be shown
\cite{matrixpencil} that the variance is given by
\begin{subequations}
\beq
{\rm var}(\omega)&=&\frac{1}{(N-L)^2 L}\frac{\sigma^2}{|h_1|^2}
\quad {\rm for} \quad L\leq N/2\,,
\\
{\rm var}(\omega)&=&\frac{1}{(N-L) L^2}\frac{\sigma^2}{|h_1|^2}
\quad {\rm for} \quad L\geq N/2\,,
\eeq
\end{subequations}
i.e. it is always {\it lower} than the corresponding variance for the KT
method, as given in Eq.~(\ref{varKT}). Choices minimizing the frequency
variance are $L=N/3$ for $L\leq N/2$, and $L=2N/3$ for $L\geq N/2$, and the
optimal frequency variance is
\be {\rm var} (\omega)=\frac{27}{4N^3}\frac{\sigma^2}{|h_1|^2}\,, \ee
as in the KT method. For one damped exponential, closed-form expressions for
the variance and bias of the damping time in the large SNR limit have recently
been derived and compared with Monte Carlo simulations \cite{djermouneKTMP}.
The results indicate that the MP method performs {\it better} (has smaller
variance) than the KT method. For both methods the frequency estimate is
unbiased, but the estimate of the damping time is biased. The bias is slightly
larger for the MP method, the difference between the two methods becoming
larger for small SNR.

\section{Fitting algorithms considered in this paper}
\label{routines}

In the rest of this paper we will compare the performance of the four fitting
algorithms described in the previous Section: Levenberg-Marquardt (LM),
modified least-squares Prony, Kumaresan-Tufts (KT) and matrix pencil (MP). For
quick reference, in Table \ref{tab:methods} we list the original papers, along
with web resources and publications providing software implementations of each
algorithm (in Fortran, MATLAB or Mathematica).  Below we give some details on
our own practical implementation of the algorithms.

\begin{table}[htb]
  \caption{Fitting methods used in this paper. El-Hadi Djermoune and
    Stanley Lawrence Marple Jr. kindly sent us up-to-date implementations of
    the codes described in \cite{djermoune} and \cite{marplebook}. We are also
    grateful to Gordon Smyth for providing us with a MATLAB routine to
    estimate parameters for purely damped exponentials, and to Boaz Porat for
    Mathematica packages to estimate parameters of undamped sinusoids using
    several methods (KT, maximum likelihood, Yule-Walker...): see
    \cite{porat}.}
  \begin{tabular}{ccc}
    Method   & Reference & Software \\
    \hline
    Levenberg-Marquardt (LM)     &\cite{LM}            &\cite{MINPACK,wolfram}\\
    Modified least-squares Prony &\cite{marplebook}    &\cite{marplebook,toolbox,fdlib,statbox}\\
    Kumaresan-Tufts (KT)         &\cite{kumaresantufts}&\cite{djermoune,mcclellan}\\
    Matrix pencil (MP)           &\cite{matrixpencil}  &\cite{djermoune}\\
  \end{tabular}
  \label{tab:methods}
\end{table}

\subsection{Levenberg-Marquardt}

The LM algorithm is more general than the other methods we consider in this
paper, in the sense that fitting functions are not restricted to a simple sum
of complex exponentials. A problem if we want to fit merger waveforms is that
LM (like most non-linear least-squares methods) is designed to deal with {\em
  real} functions. For this reason the authors of \cite{Buonanno:2006ui}, who
studied waveforms similar to those in Fig.~\ref{wf1}, fitted only the real (or
imaginary) part of the signal.  Some practice also shows that general
non-linear least-squares methods (such as the LM algorithm) often fail to
converge if the signal is given by a superposition of damped sinusoids, unless
we provide very accurate initial guesses for the fitting parameters.

For simplicity, and to take into account these limitations of non-linear
least-squares algorithms, in Section \ref{fit-performance} we will compare the
performance of different routines on a {\em real} ringdown signal with a {\em
  single} frequency and damping time.  We choose some starting time $t_0$ and
we fit the real signal by a four-parameter function:
\be\label{fitfunc}
\Psi(t)=A e^{\alpha (t-t_0)} \sin(\omega t+\varphi)\,,
\ee
The LM algorithm provides reasonable answers only if the four parameters
$(A,~\varphi,~\omega,~\alpha)$ are reasonably close to their ``true'' values.
In particular, an accurate initial guess for $A$ is necessary to avoid ``hard
failures'' of the fitting routine: we choose as initial guess the value of the
signal amplitude at $t_0$. We fix the tolerance parameter in the \url{lmdif}
routine to be \url{TOL}$=10^{-12}$ (but this choice is not crucial).

As a final remark we point out that, by using a real signal, in a sense we are
``biasing'' our tests in favour of the LM method.  The reason is that fitting
a real signal requires the inclusion of at least {\em two} complex
exponentials in the sum (\ref{signal}), unnecessarily doubling the number of
unknown parameters to be searched for by Prony methods.

\subsection{Modified least-squares Prony, Kumaresan-Tufts and matrix pencil}

A Fortran routine implementing modified least-squares Prony was kindly
provided to us by Stanley Lawrence Marple Jr. \cite{marplebook}. We tested the
routine's performance on noiseless ``pure ringdown'' waveforms obtained by
superposing three damped exponentials with frequencies given by the first
three QNMs $(n=0,~1,~2)$ of a Schwarzschild black hole \cite{bcw}. For
noiseless waveforms, the frequency and damping time of the fundamental mode
($n=0$) are typically determined with accuracies $\sim 10^{-6}$. The frequency
and damping time of the two overtones usually have accuracies $\sim 10^{-5}$.
Parameter estimation becomes much worse, as expected from the theory, when we
add even a modest amount of noise to the waveforms.

Performance in the presence of noise gets much better when we use two MATLAB
routines by El-Hadi Djermoune \cite{djermoune}, implementing the
Kumaresan-Tufts and matrix pencil methods.  These routines require the
specification of the prediction order (that we always choose to be $L=N/3$,
where $N$ is the number of data points, in order to minimize the variance in
parameter estimation). In addition, they also require the specification of the
number of complex exponentials to be searched for.  Djermoune kindly provided
us with an additional routine to {\it estimate} the number of complex
exponentials present in the signal, or (in more technical jargon) the ``order
of the exponential model''.  The estimation routine is based on two criteria:
Akaike's information criterion (AIC) and the minimum description length (MDL)
criterion \cite{reddybiradar,waxkailath}.

When tested on noiseless waveforms, Djermoune's routines are remarkably
successful at estimating the number of modes present in the signal.
Frequencies and damping times for noiseless signals are typically determined
to accuracies $\sim 10^{-15}$ (comparable to machine precision). 

We will see in the following that MP and KT methods have essentially the same
performance. In fact, in the next Section we will show by numerical
experiments that MP methods have slightly smaller variance and bias in
parameter estimation: this was pointed out also by Djermoune and collaborators
(see eg.  \cite{djermoune,djermouneKTMP,djermouneSNR}). Due to the similarity
of the two techniques, and due to the slightly superior performance of MP over
KT, we will usually compare MP and LM methods in our analysis of merger
waveforms (Section \ref{fitmerger} below).

As a final remark, we point out that MP and KT have a variance in frequency
estimation which is only $\simeq 9/8=1.125$ times larger than the Cramer-Rao
bound.  These algorithms may prove extremely useful not only to study the
merger waveforms produced by numerical relativity (as we do here), but also to
estimate the source parameters after GW detection. MP and KT algorithms are
most effective and useful when we deal with noisy waveforms, and these
situations are often the most interesting. For example, fitting the late-time,
low-amplitude portion of a numerical relativity waveform yields the remnant
black hole's parameters. Fitting large-$l$ multipolar components (that carry
little energy, hence are more affected by numerical noise) is necessary for
tests of the general relativistic no-hair theorem \cite{bcw}. Last but not
least, the importance of parameter estimation from experimental, noisy GW data
can hardly be underestimated.

\section{Comparison of different methods on noisy damped sinusoids}
\label{fit-performance}

To test the performance of different fitting methods, we built model waveforms
reproducing the essential features of the ringdown waveforms produced by
binary black hole merger simulations (see Fig.~\ref{wf1}). For simplicity,
and to take into account the limitations of the LM algorithm, here we consider
our ``true'' signal as consisting of a single damped sinusoid. We take the
frequency and damping time to be those of a the fundamental ($n=0$) $l=m=2$
perturbations of a Kerr black hole, as listed in \cite{bcw}. For a given
oscillation frequency $\omega$, we produce a signal lasting five GW cycles:
i.e. the signal length $t_{\rm fin}=5(2\pi/\omega)=5T_{\rm GW}$. The sampling
time is taken to be $T=T_{\rm GW}/50$, so that the ``full'' waveform consists
of 250 data points.

To this ``true'' signal we superimpose Gaussian white noise. In this Section
we do not consider standard least-squares Prony, since it is outperformed by
MP and KT algorithms in the presence of noise. The MP and KT algorithms are
implemented in MATLAB. In this case, we produce Gaussian noise using the
built-in function \url{normrnd}.  For our Fortran implementation of the LM
algorithm we generate noise as a random variable $\varpi$ with mean $\mu=0$
and standard deviation $\sigma$ using the Box-M\"uller method
\cite{boxmuller}: for each $t$, given two random numbers $u_1$ and $u_2$
uniformly distributed in $[0,~1]$ (as generated with the Numerical Recipes
routine \url{ran2} \cite{numrec}), we add to our signal
\be
\varpi(t)=\mu+\sigma\sqrt{-2\ln(u_1)}\cos(2\pi u_2)\,.
\ee
\begin{figure*}[ht]
\begin{center}
\epsfig{file=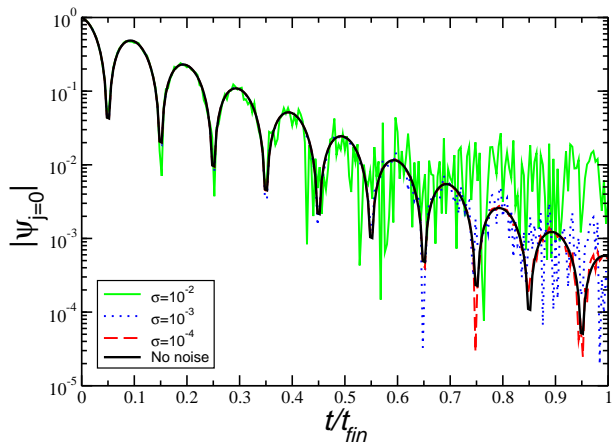,width=7cm,angle=-90} 
\end{center}
\caption{QNM waveforms with and without Gaussian white noise. We normalize the
  time axis to the total duration of the signal $t_{\rm fin}$. The ``pure''
  noiseless waveform (thick black line) has unit amplitude.  Dashed (red),
  dotted (blue) and thin (green) lines are the same waveform superimposed to
  Gaussian white noise with $\sigma=10^{-4},~10^{-3},~10^{-2}$, respectively.}
\label{fig:data}
\end{figure*}

For illustration, in Fig.~\ref{fig:data} we show a typical waveform generated
in this way. 
The ``true'' signal (with $\sigma=0$) corresponds to a perturbed Schwarzschild
black hole, i.e.  the frequency and damping time correspond to the fundamental
$l=m=2$ QNM for a black hole of dimensionless spin parameter $j=0$ \cite{bcw}.
For any given $\sigma$, the addition of Gaussian white noise has a less severe
impact for large values of $j$. The reason is that the damping time of the
$l=m=2$ mode grows with the rotation parameter $j$ (see eg.  Table II in
\cite{bcw}). A Schwarzschild waveform with $\sigma=10^{-3}$ is quite similar
to the typical merger waveforms of Fig.~\ref{wf1}, and for this reason we will
use it to test the performance of the different fitting routines.  Results for
different values of $\sigma$ are qualitatively similar.

For each fitting algorithm, we perform Monte Carlo simulations to produce a
large number ($N_{\rm noise}=100$) of realizations of the noise with
$\sigma=10^{-3}$. For each realization of the noise ($\nu=1,\dots,N_{\rm
  noise}$) we fit the waveform varying the starting time from $t_0=0$ to
$t_0=t_{\rm fin}$. Each fit yields a couple of values
$\omega_{\nu}(t_0),~\alpha_{\nu}(t_0)$, and from these we can deduce the
quality factor of the oscillation $Q_{\nu}(t_0)\equiv
|\omega_\nu(t_0)/[2\alpha_\nu(t_0)]|$ \cite{bcw}. For each $t_0$ we compute
the standard deviation and bias of each of these fitted quantities $x$
($=\omega,~\alpha$ or $Q$) from our Monte Carlo simulations:
\beq 
\left.{\rm std}(x)\right|_{t_0}&=&
\left\{
\f{1}{{N_{\rm noise}-1}}
\sum_{\nu=1}^{N_{\rm noise}} [x_{\nu}(t_0)-\bar x(t_0)]^2
\right\}^{1/2}
\,,\\
\left.{\rm bias}(x)\right|_{t_0}&=&\bar x(t_0)-x_{\rm true}\,,
\eeq
where $x_{\rm true}$ is the known, true value of each quantity and a bar
denotes the average over all noise realizations:
\be
\bar x(t_0)
=\left[\f{1}{N_{\rm noise}}
\sum_{\nu=1}^{N_{\rm noise}} x_{\nu}(t_0)\right]\,.
\ee
\begin{figure*}[ht]
\begin{center}
\begin{tabular}{cc}
\epsfig{file=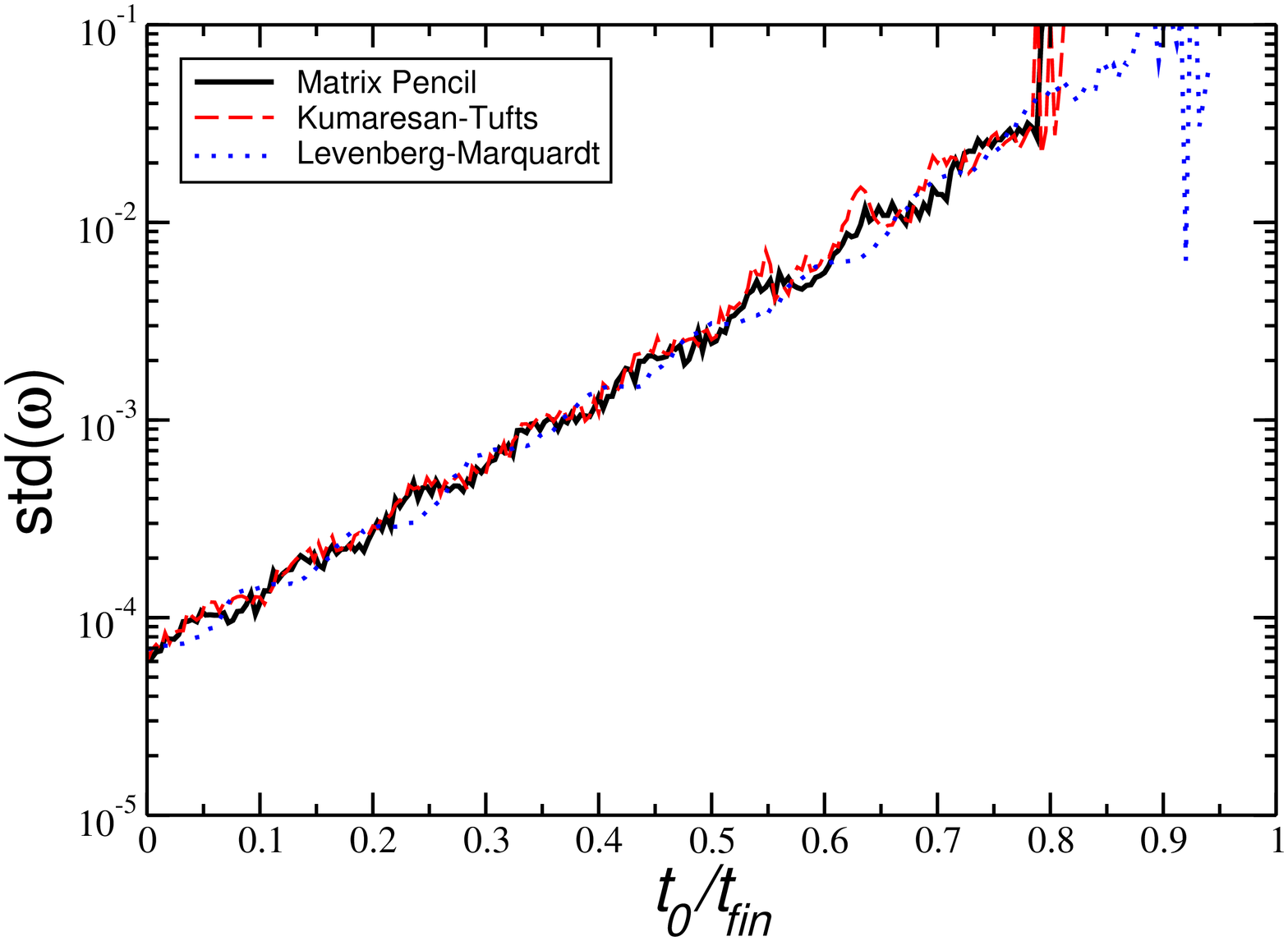,width=7cm,angle=270} &
\epsfig{file=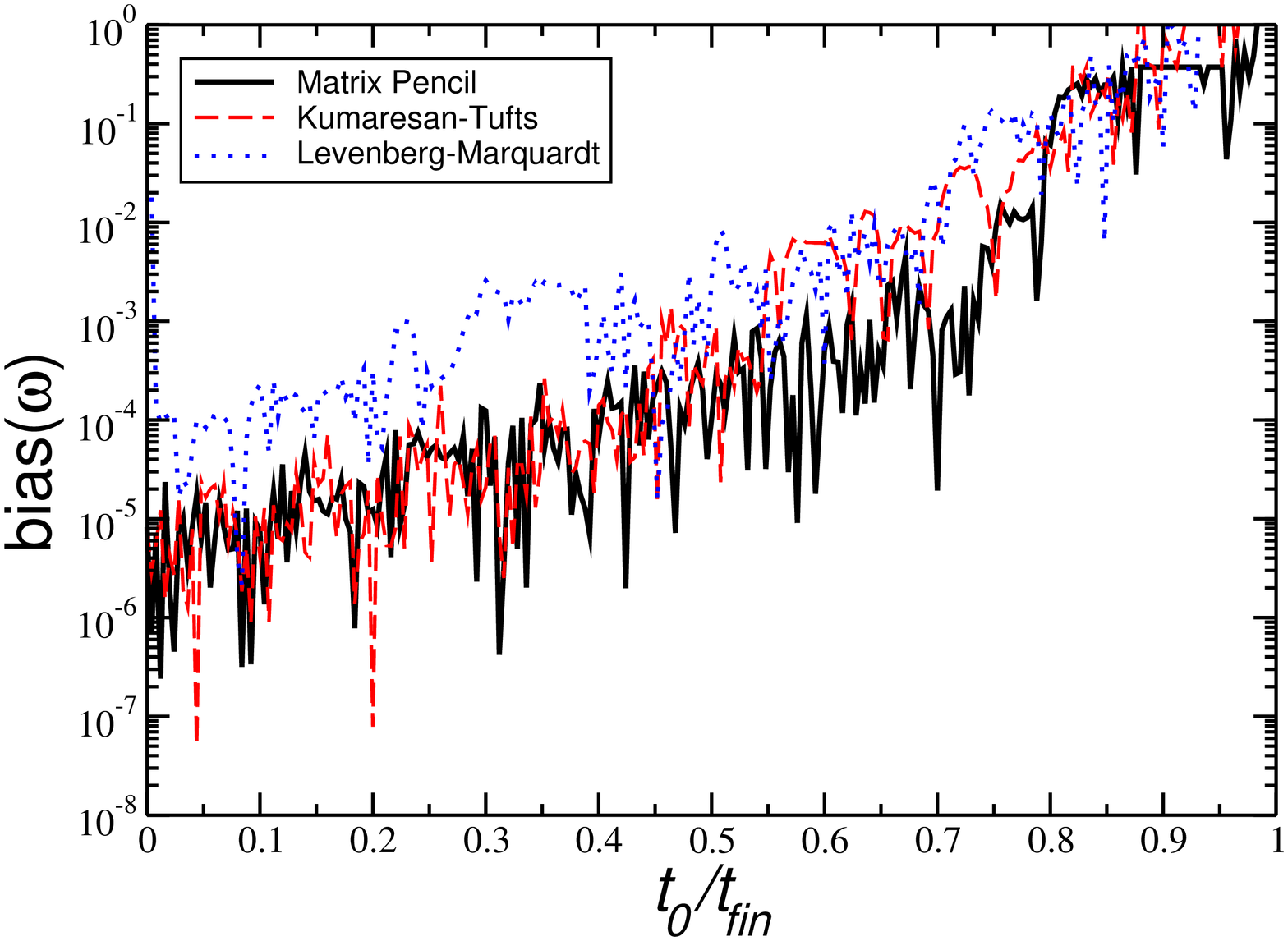,width=7cm,angle=270} \\
\epsfig{file=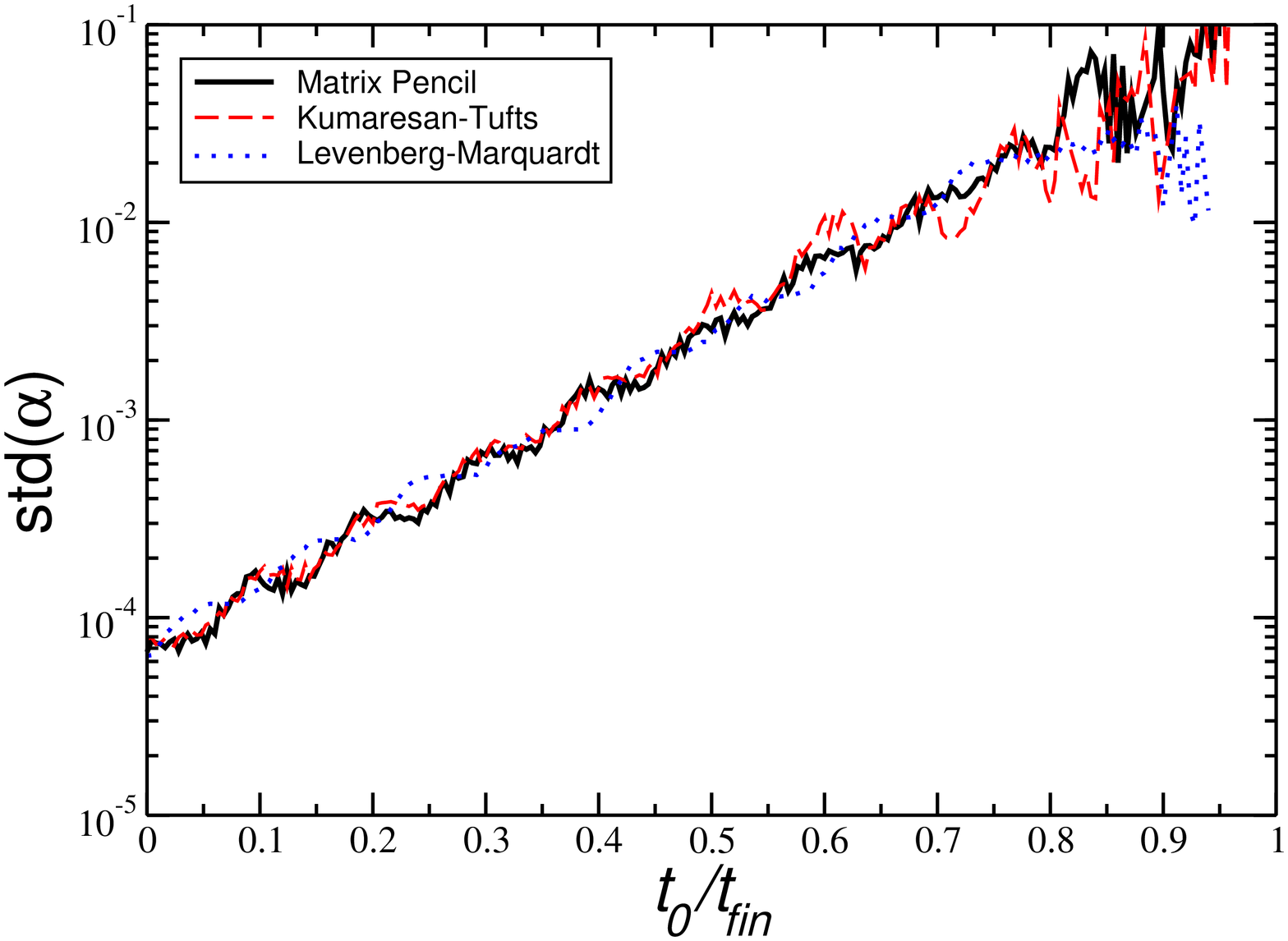,width=7cm,angle=270} &
\epsfig{file=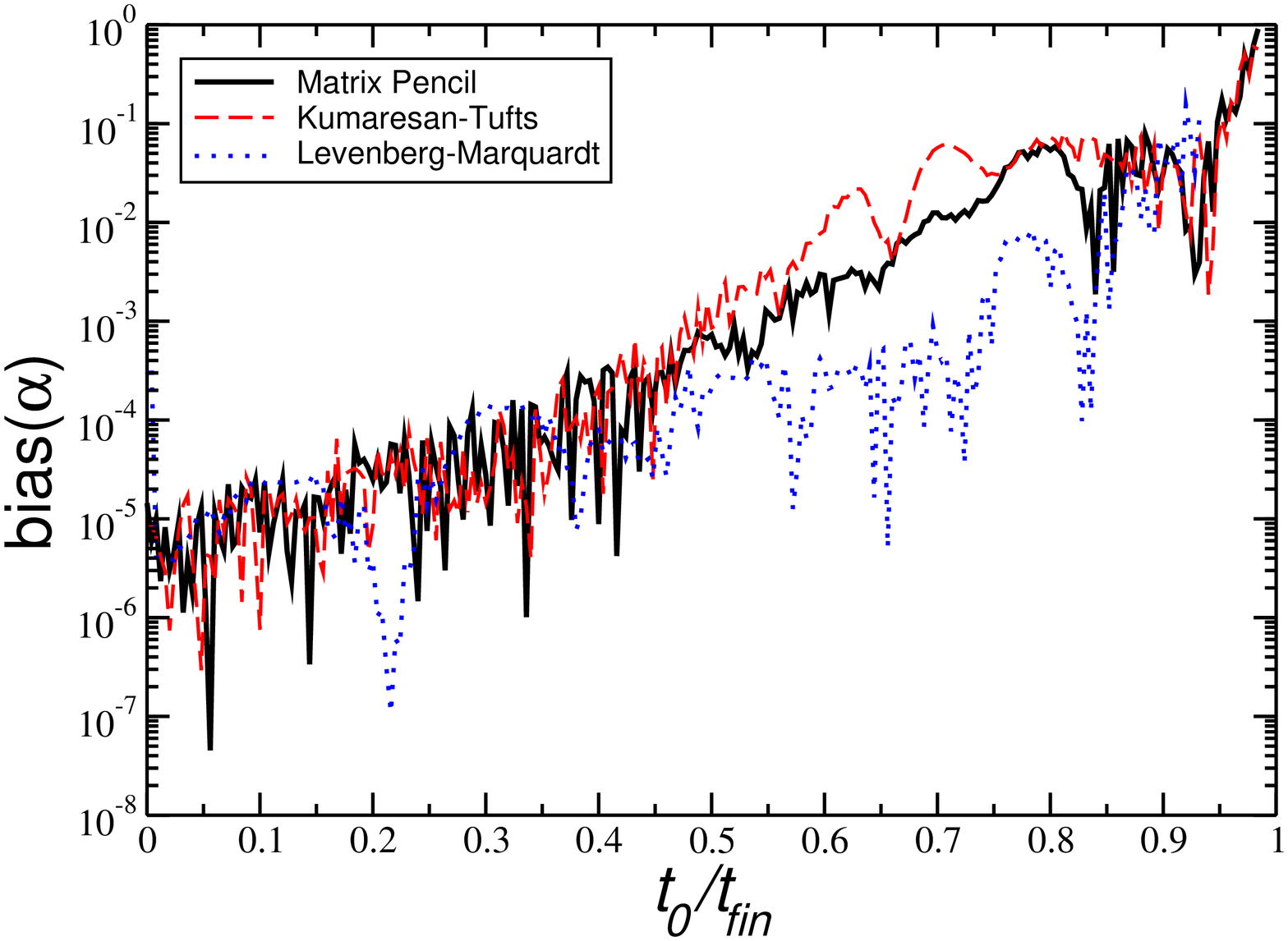,width=7cm,angle=270} \\
\epsfig{file=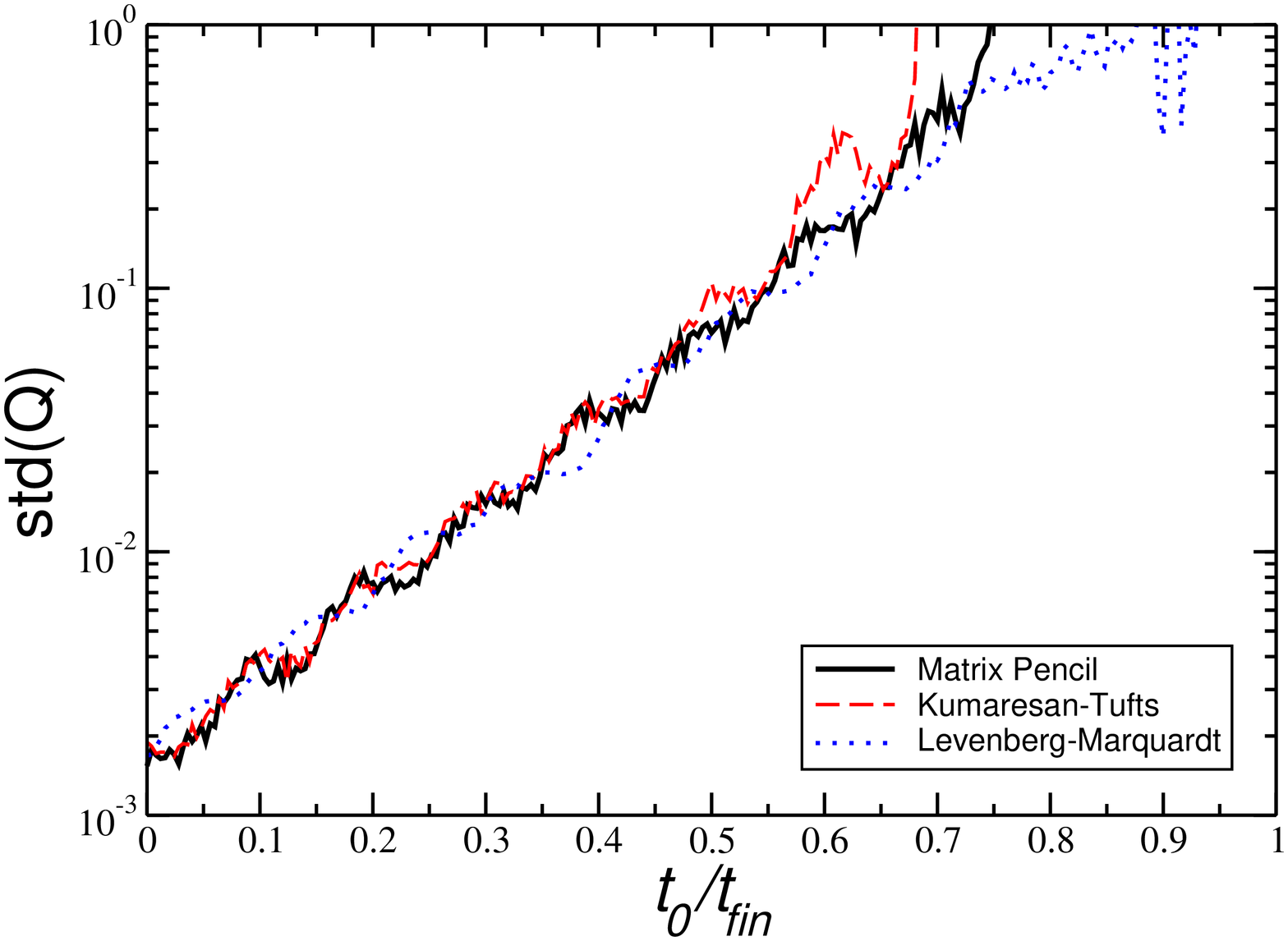,width=7cm,angle=270} &
\epsfig{file=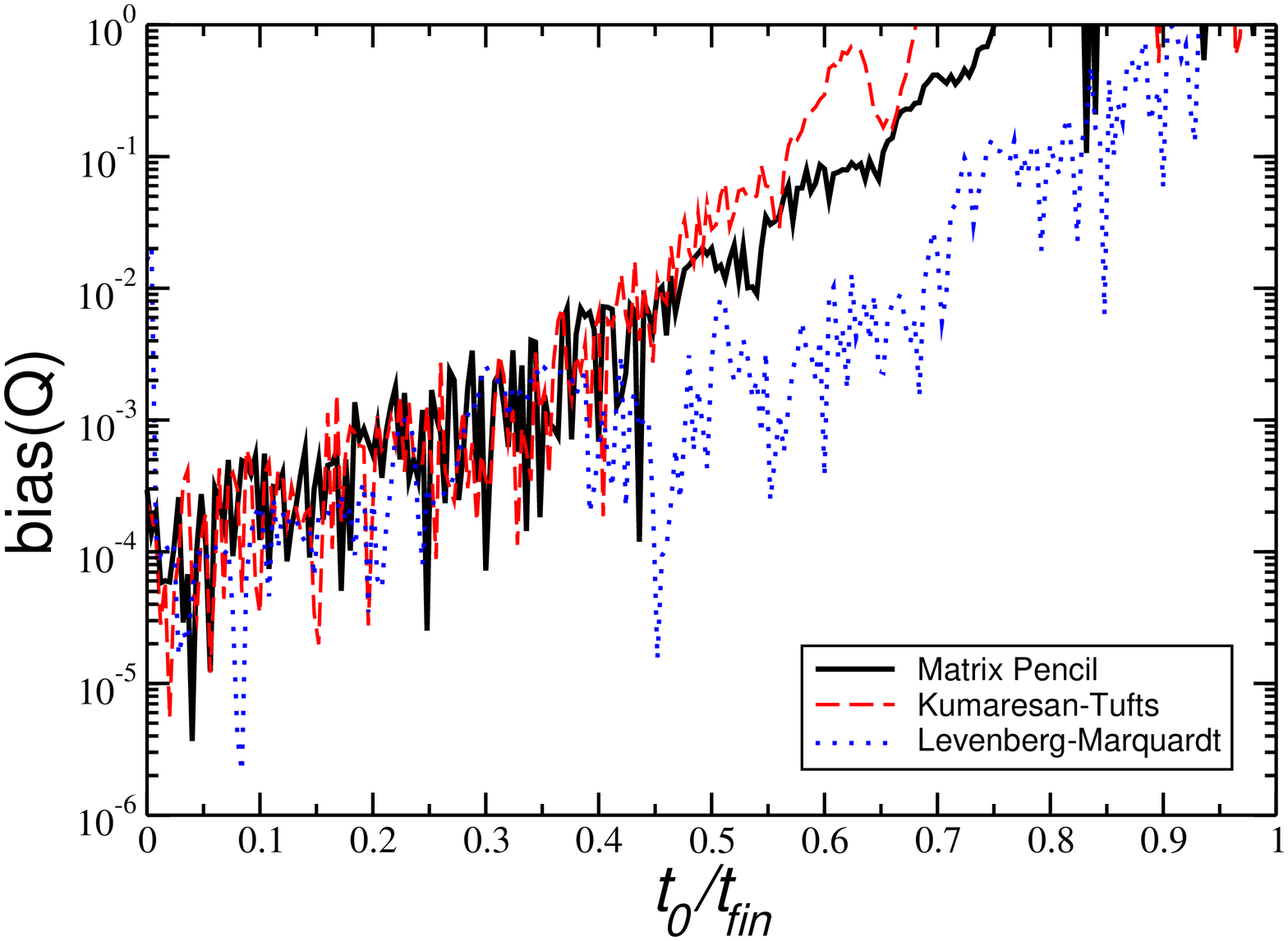,width=7cm,angle=270} \\
\end{tabular}
\end{center}
\caption{Standard deviation (left) and bias (right) in the estimate of
  frequency, damping time and quality factor (top to bottom). All quantities
  are given as functions of the starting time of the fit $t_0$ (normalized by
  the duration of the signal $t_{\rm fin}$); each point is the result of a
  Monte Carlo simulation obtained by adding $N_{\rm noise}=100$ realizations
  of Gaussian white noise with zero mean and $\sigma=10^{-3}$ to the $j=0$
  waveform of Fig.~\ref{fig:data}.  Solid (black), dashed (red) and dotted
  (blue) lines refer to the MP, KT and LM algorithms, respectively.}
\label{fig:var-bias}
\end{figure*}

Results of these Monte Carlo simulations are shown in Fig.~\ref{fig:var-bias}.
There we see that all three methods (KT, MP and LM) are essentally equivalent
in terms of variance\footnote{We also tested the MP and KT methods on pure
  sinusoids, finding good agreement with theoretical predictions for the
  frequency variance.}. Standard deviation and bias grow sharply for
$t_0/t_{\rm fin}\gtrsim 0.7$, when we are fitting that part of the signal
which is buried in noise (compare Fig.~\ref{fig:data}). The bias is usually
very small, but remarkably linear methods (KT, and especially MP) beat the
non-linear LM fitting routine in terms of bias on the frequency. In terms of
bias on the damping time, LM performs slightly better than MP and KT only for
large values of $t_0$, when the SNR is very small.

When comparing the performance of the different methods it is useful to
remember that LM only works if we give a good initial guess for the parameters
we want to estimate, whereas MP and KT {\it automatically} find the values of
these parameters, with no need for initial guesses. Our comparison is somehow
``biased'' in favour of the LM method in at least three ways: (i) we choose
optimal initial guesses for the parameters, so that the LM algorithm is not
``allowed'' to converge to some wrong root; (ii) we choose a real signal
rather than a complex signal, because the LM algorithm only works for real
data sets; (iii) we only consider one damped exponential (LM often fails to
converge if we include additional damped sinusoids, whereas Prony-type methods
are still remarkably successful).

\section{Application to merger waveforms}
\label{fitmerger}

Here we turn to the problem that motivates the present analysis of fitting
algorithms for complex exponentials in noise. If the ringdown radiation
emitted as a result of a binary merger shows no signs of non-linearities or
mode coupling, our fits should have a simple behavior: as we increase the
starting time for the fit $t_0$, the black hole's oscillation frequencies
$\omega_{lmn}$ and damping factors $\alpha_{lmn}$ should {\em asymptotically
  approach a constant}, whose value is predicted by linear perturbation theory
\cite{bcw}.

\begin{figure*}[ht]
\begin{center}
\begin{tabular}{cc}
\epsfig{file=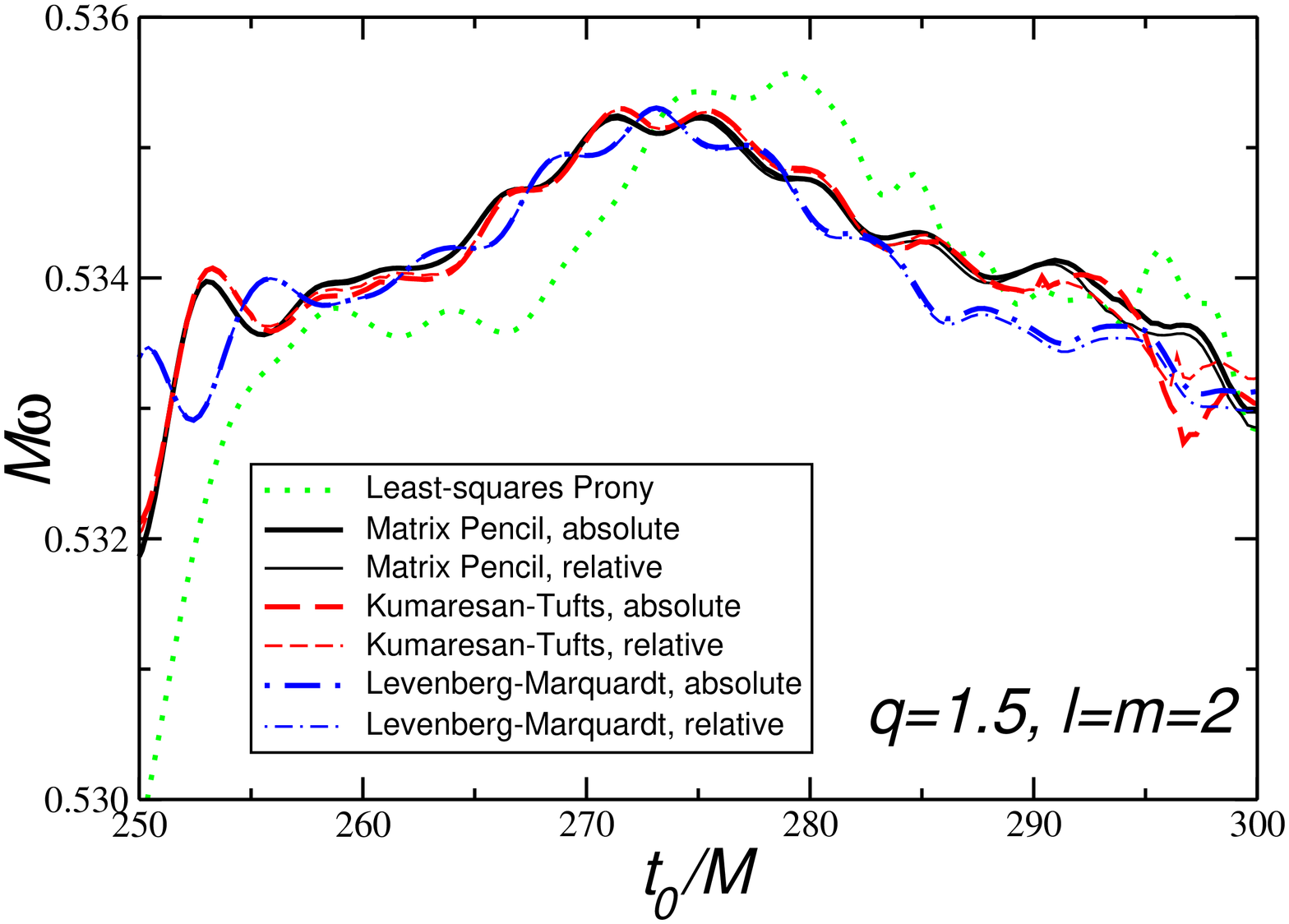,width=7cm,angle=-90}&
\epsfig{file=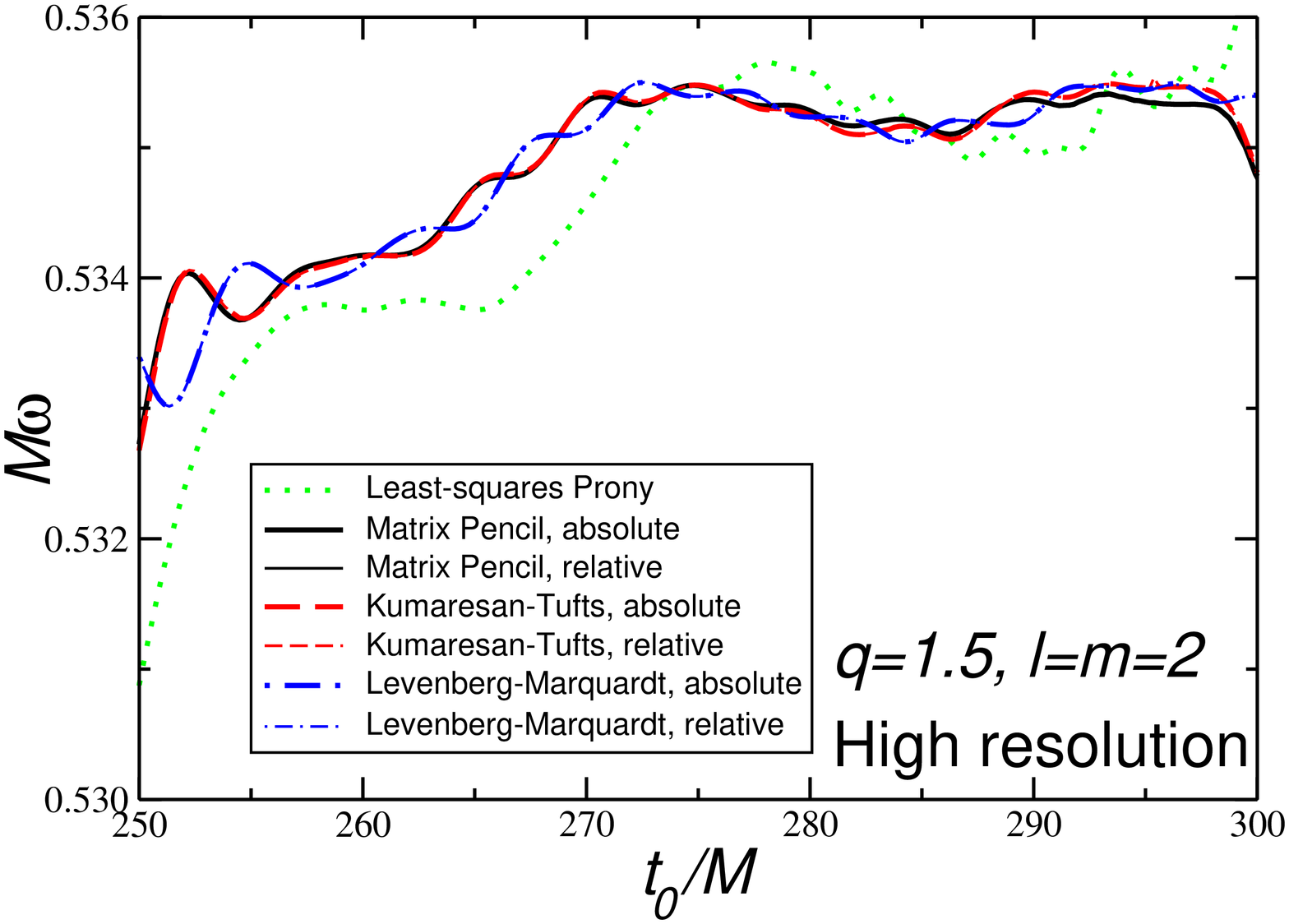,width=7cm,angle=-90}\\
\epsfig{file=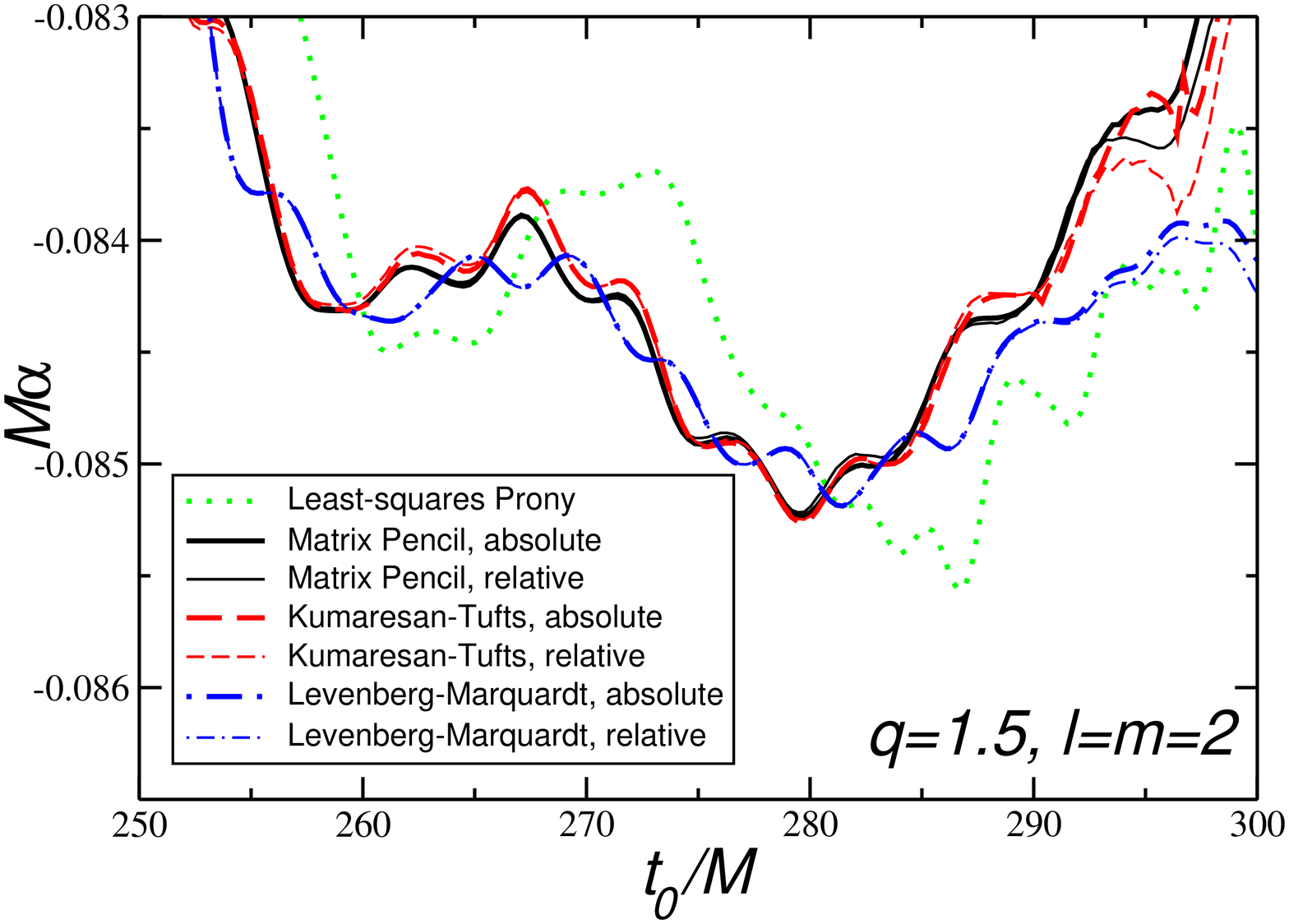,width=7cm,angle=-90}&
\epsfig{file=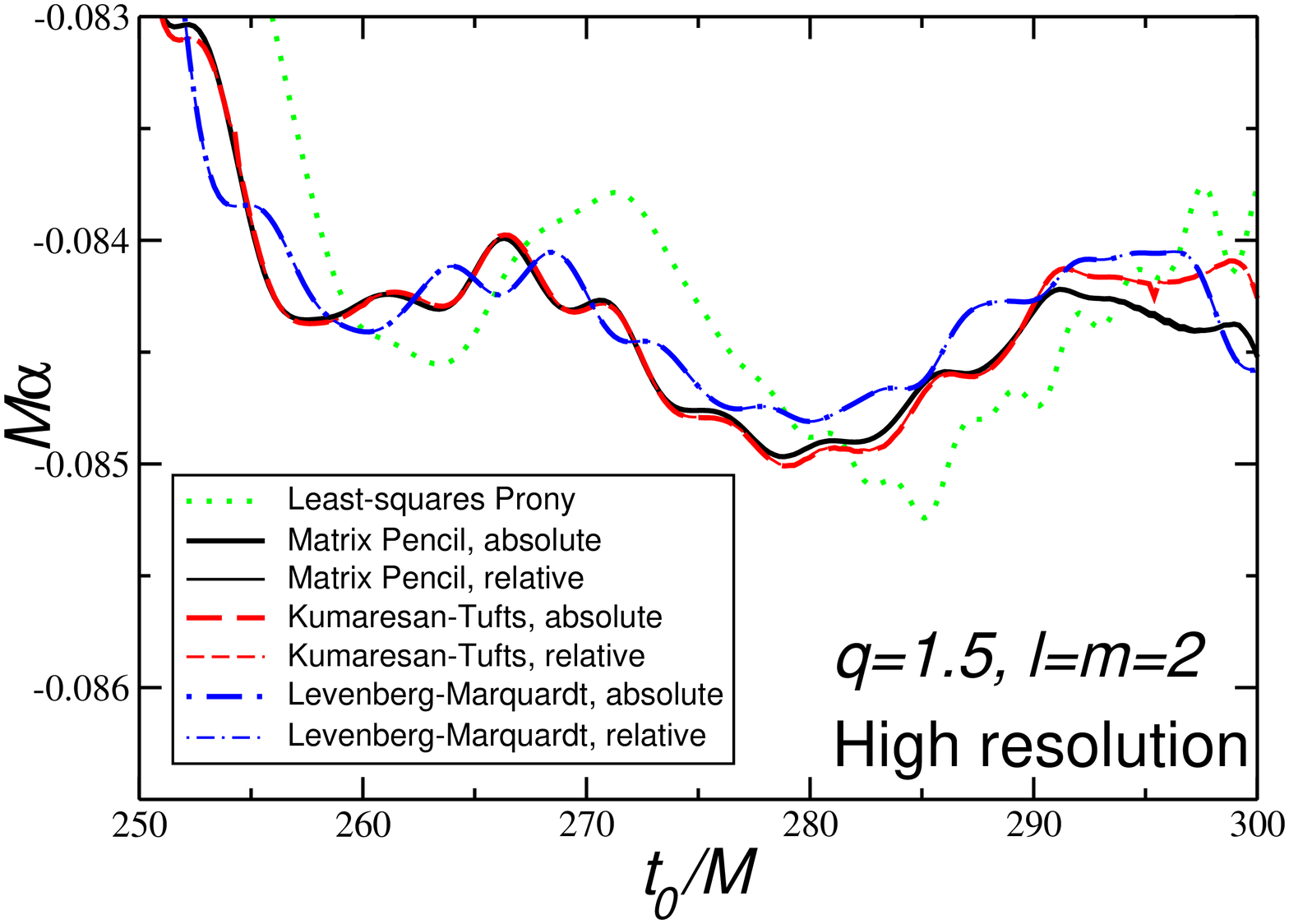,width=7cm,angle=-90}\\
\end{tabular}
\caption{Performance of different Prony methods in estimating the oscillation
  frequency $\omega$ (top) and damping factor $\alpha$ (bottom) for low and
  high-resolution runs (left and right, respectively). For concreteness we
  choose a binary with $q=1.5$ and consider the fundamental mode with $l=m=2$.
  Prony-type results refer to the complex waveform, while LM results are
  obtained by fitting the {\em real} part of the waveform only.
  \label{performance}}
\end{center}
\end{figure*}

\begin{figure*}[ht]
\begin{center}
\begin{tabular}{cc}
\epsfig{file=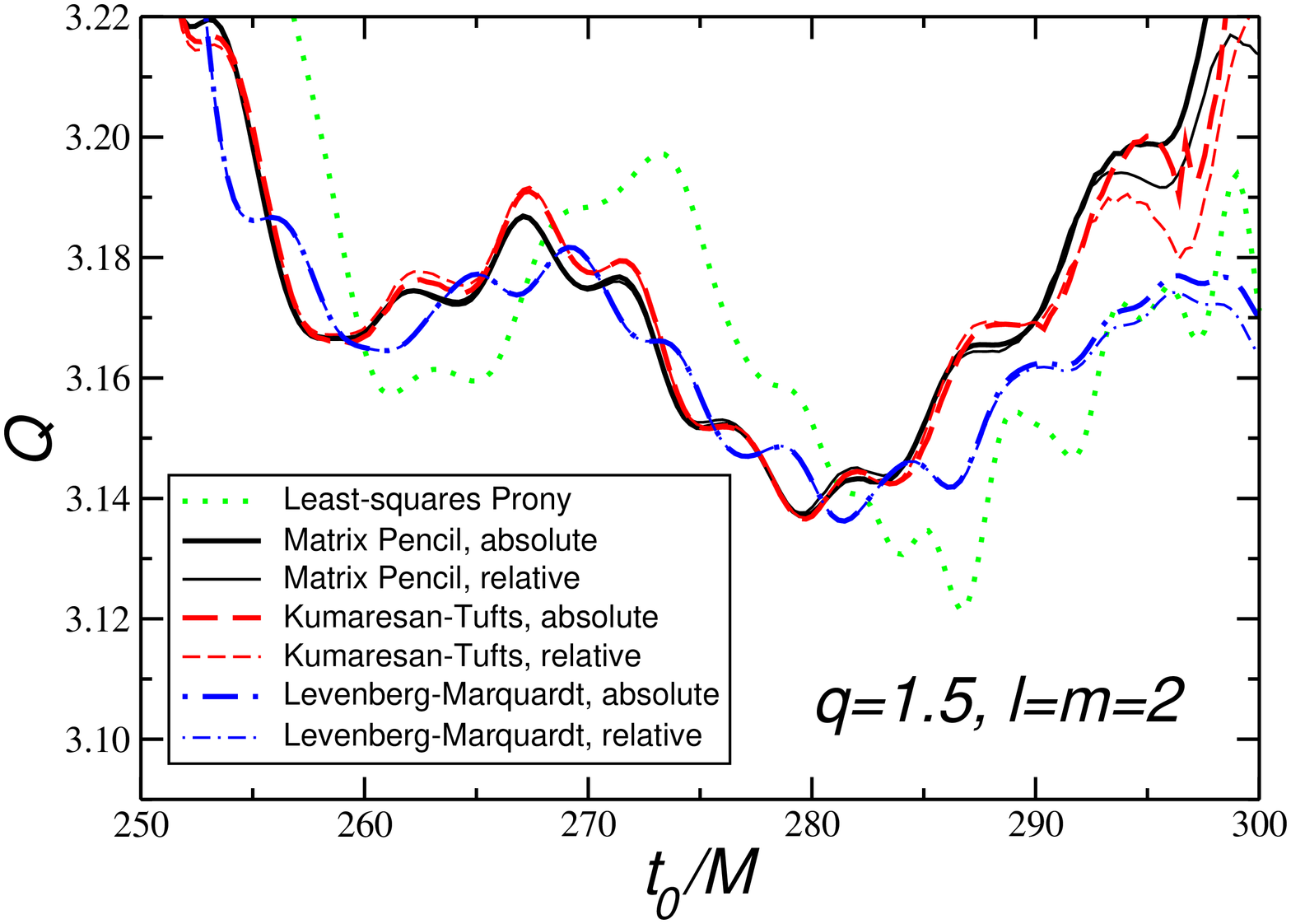,width=7cm,angle=-90}&
\epsfig{file=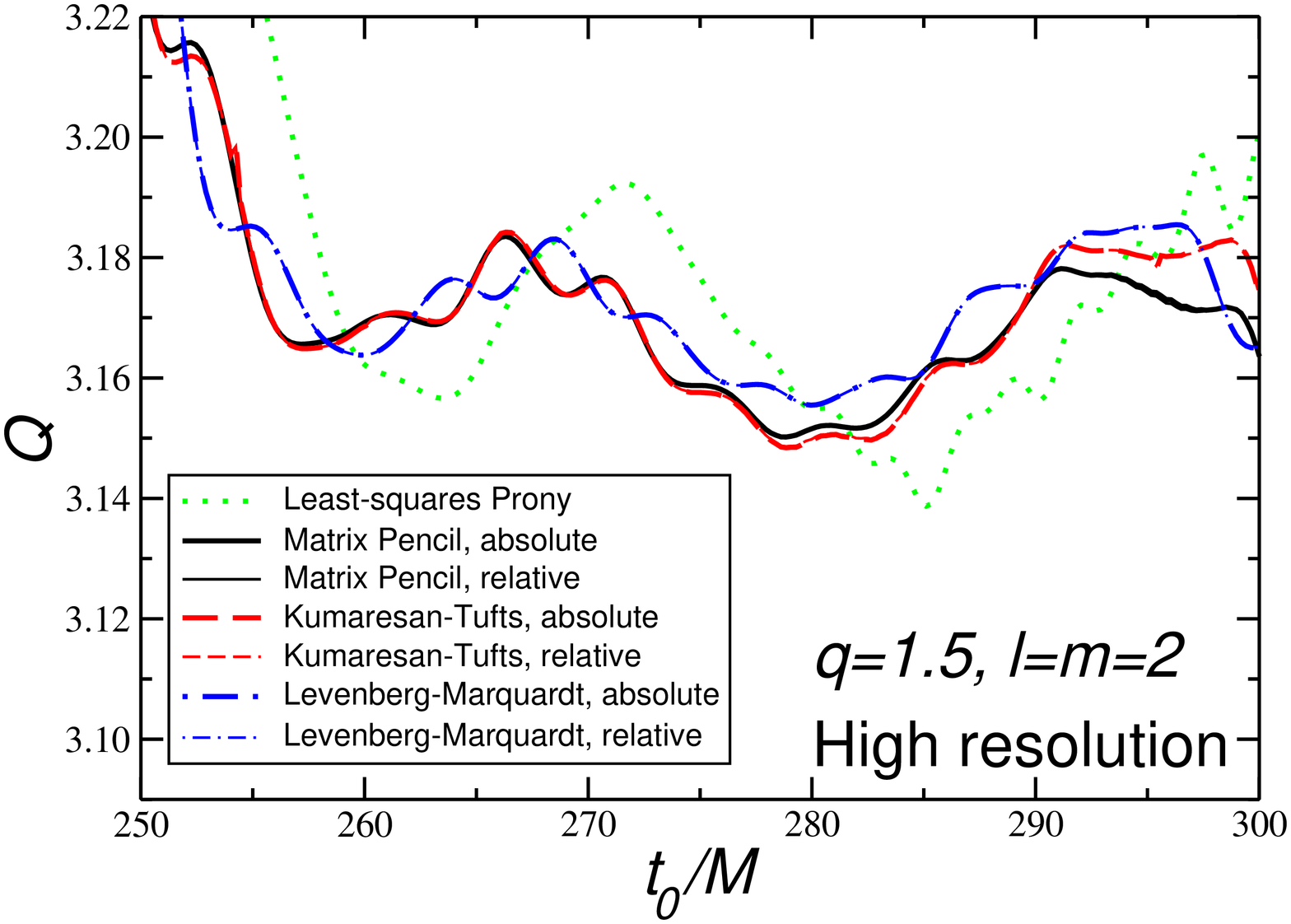,width=7cm,angle=-90}\\
\end{tabular}
\caption{Same as Fig.~\ref{performance}, but for the quality factor
  $Q$.
  \label{performanceq}}
\end{center}
\end{figure*}

In Fig.~\ref{performance} we show results of QNM fits performed on the $l=m=2$
component of low-resolution and high-resolution merger simulations with mass
ratio $q=1.5$. Solid lines refer to the ``absolute'' truncation criterion of
Section \ref{waves}, and dashed lines to the ``relative'' truncation
criterion. Different truncation criteria affect the estimated parameters only
for low-resolution simulations, and at late starting times ($t_0/M\gtrsim
290$).  Results obtained fitting by a single complex exponential with standard
least-squares Prony\footnote{Figs.~\ref{performance} and \ref{performanceq}
  only display results for least-squares Prony obtained using the ``relative''
  truncation criterion. However, we did check that using the ``absolute''
  truncation criterion has a very small effect on the estimated parameters.
  We also checked that increasing the number of exponentials we search for
  does {\em not} improve the agreement of standard least-squares Prony with
  other methods.}  are slightly off from the ``best'' fitting methods (MP, KT
and LM). Not surprisingly, there is remarkable agreement between KT and MP
methods, and very good agreement between these two methods and the non-linear
least-squares LM fit.

The main conclusion to be drawn from Fig.~\ref{performance} is that {\em all
  fitting routines consistently predict that frequencies and damping times
  have small (but non-trivial in structure) time variations}. Relative
variations in $\omega$ are of order $\sim 0.5\%$, with the frequency reaching
a local maximum at $t_0/M\sim 272$. Relative variations in $\alpha$ are
slightly larger (roughly $\sim 2.5\%$), with $|\alpha|$ attaining a maximum
(and the damping time correspondingly attaining a minimum) for $t_0/M\sim
280$.  Notice also that increasing the resolution sensibly reduces the
irregularities in the predicted frequency at all times, and produces a
flattening of the estimated parameters for $280\lesssim t_0/M\lesssim 300$.
The decrease in $\omega$ and in $|\alpha|$ for $280\lesssim t_0/M\lesssim
300$, that are visible in the left panels, are clearly an artifact of
insufficient resolution. It would be interesting to perform Richardson
extrapolation of the numerical results to determine if oscillations in the
estimated parameters (which could be a sign of ``new'' physics) disappear in
the limit of infinite resolution. We plan to address this problem in the near
future. An analysis of the fine structure of the signal, and of its
implications for gravitational wave phenomenology, will be presented in a
forthcoming publication \cite{follow-ups}.

For the time being, we simply point out that such systematic variations in the
oscillation frequency and/or damping time could be signs of non-linearities
and/or mode coupling in the numerical simulations. Variations could be due to
the black hole's mass and angular momentum changing on a timescale which is
longer than the QNM oscillation period, or to beating phenomena with other QNM
frequencies. In our opinion, the fact that all fitting methods consistently
predict the same ``global'' structure is convincing evidence that the
existence of maxima and minima is {\it not} due to numerical errors in the
fit.

In Fig.~\ref{performanceq} we plot the quality factor of the oscillations as a
function of $t_0/M$. The time variation of $Q$ is basically dominated by the
time variation of $\alpha$. This is quite obvious, since $Q=\omega/(2\alpha)$
and time variations in $\alpha$ are larger than time variations in $\omega$.
The interest of plotting $Q(t_0/M)$ comes from the fact that, in linear theory
and for $l=m$, $Q$ is a monotonic function of $j$ \cite{bcw}. Roughly
speaking, large variations in $Q$ for the fundamental oscillation mode could
mean that the Kerr angular momentum parameter is changing in time, or that
there are significant corrections to the linear approximation.

\begin{figure*}[ht]
\begin{center}
\begin{tabular}{cc}
\epsfig{file=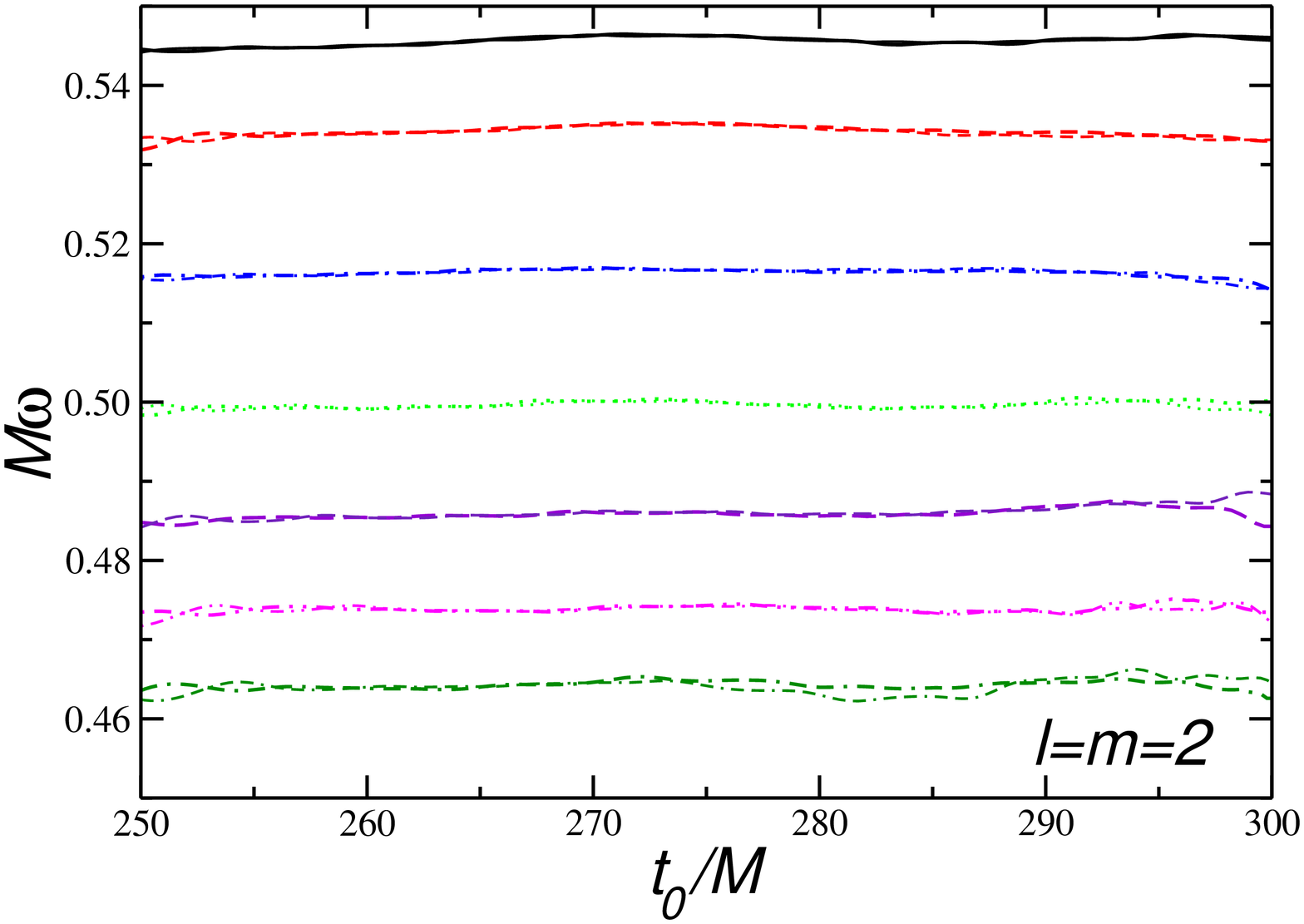,width=7cm,angle=-90}&
\epsfig{file=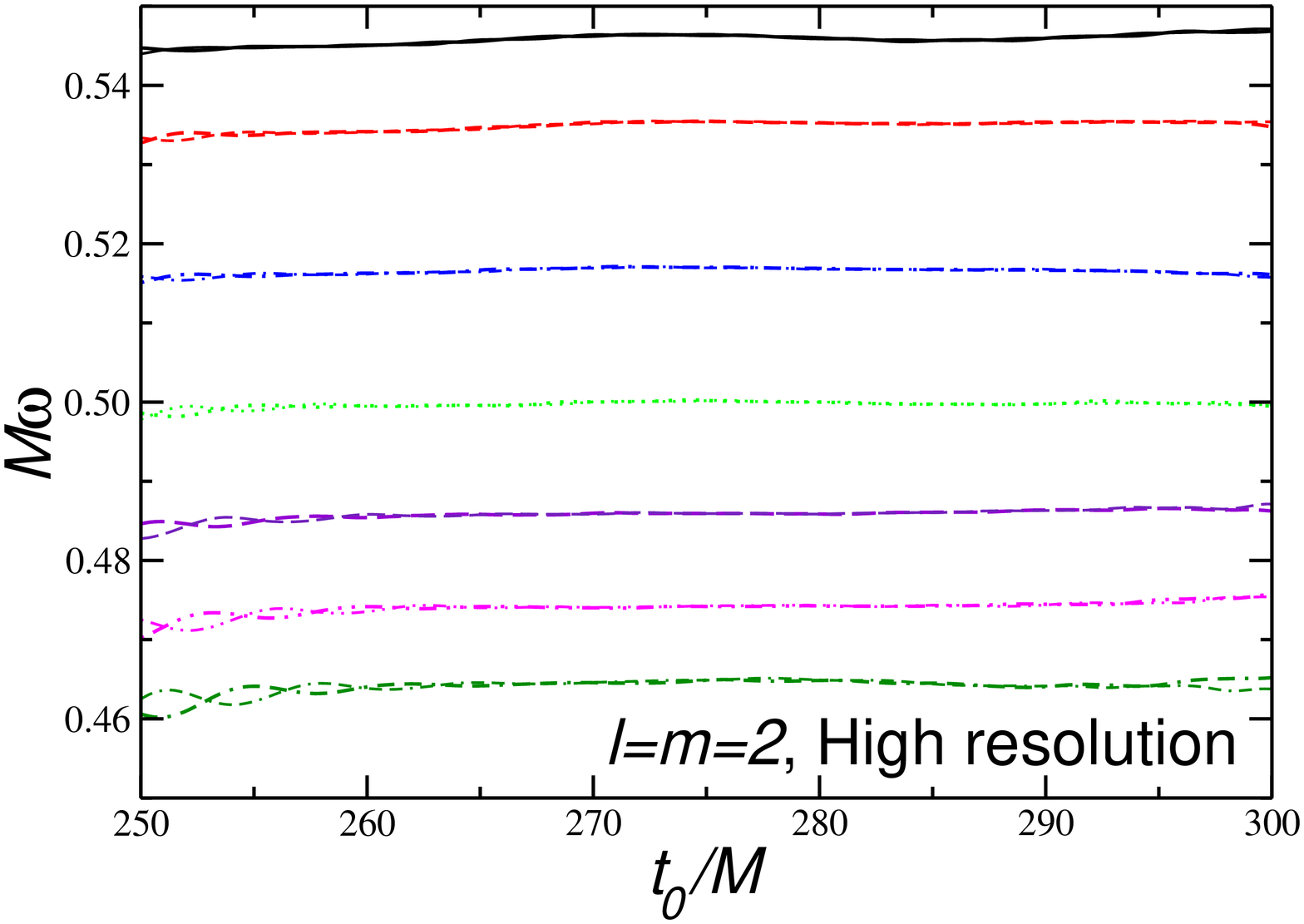,width=7cm,angle=-90}\\
\epsfig{file=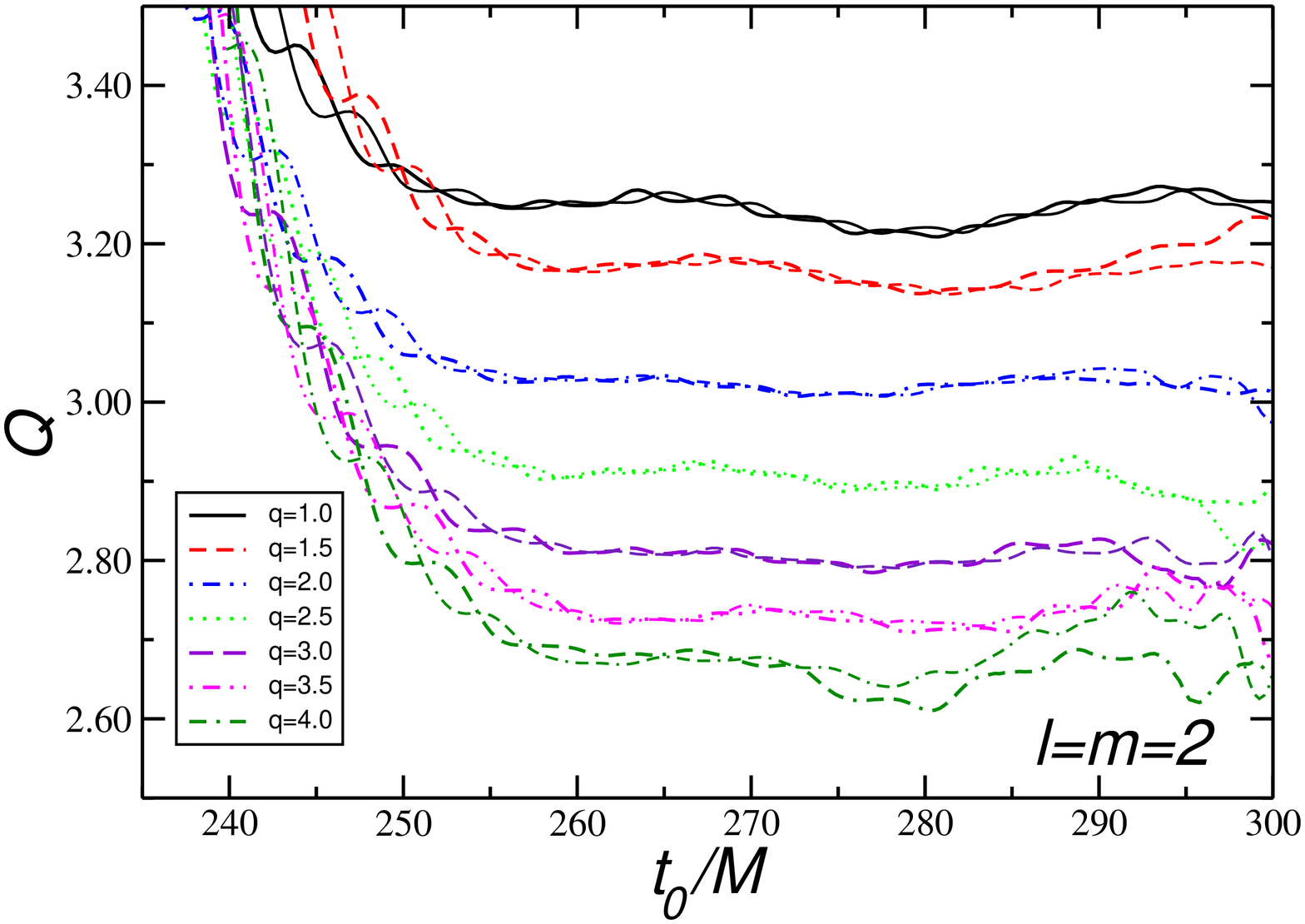,width=7cm,angle=-90}&
\epsfig{file=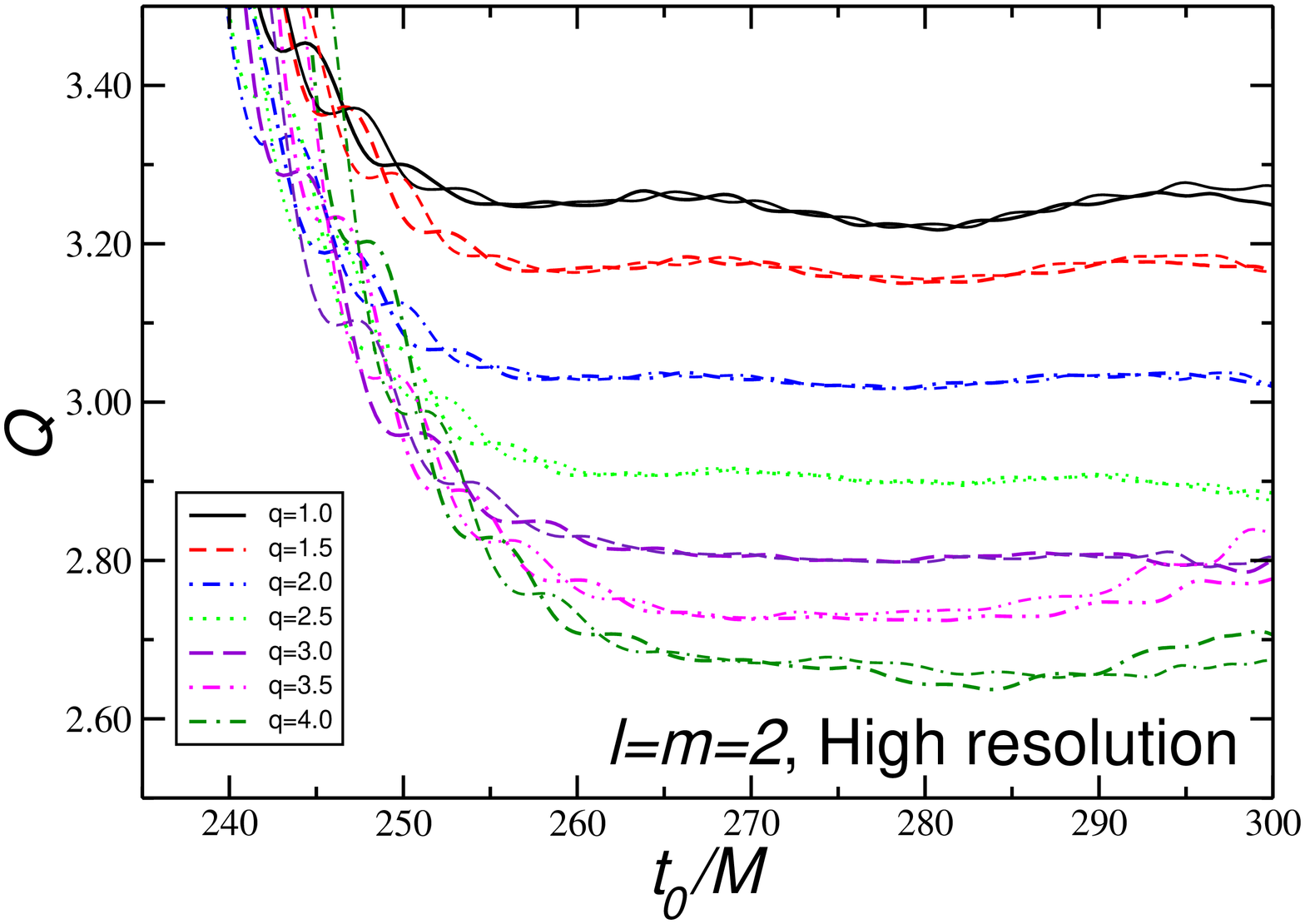,width=7cm,angle=-90}\\
\end{tabular}
\caption{Oscillation frequency (top) and quality factor (bottom) for modes
  with $l=m=2$ and for low and high-resolution runs (left and right,
  respectively), as a function of the starting time of the fit. Thick lines
  use MP, thin lines use LM. Values of the quality factor for $t/M\lesssim
  250$ are often unphysically large. We truncate the waveform at late times
  when the absolute value of the amplitude drops below $10^{-4}$ (for low
  resolution) or $10^{-5}$ (for high resolution). In each panel, lines from
  top to bottom refer to $q=1.0,~1.5,~2.0,~2.5,~3.0,~3.5,~4.0$.
  \label{wrQ}}
\end{center}
\end{figure*}

Finally, in Fig.~\ref{wrQ} we compare the performance of our two ``best''
fitting methods (MP and LM) in estimating oscillation frequencies and quality
factors for different mass ratios: $q=1.0,~1.5,~2.0,~2.5,~3.0,~3.5,~4.0$.
Deviations in $Q$ between the two methods are quite significant, especially
for large mass ratio, when the numerical simulations are less reliable. For
$q=4.0$ and low resolution, the LM method shows larger variations in $\omega$
than the MP method. This could be an effect of the larger bias in frequency of
the LM method, as compared with the MP method (cf.  Fig.~\ref{fig:var-bias}).
Once again, increasing the resolution produces a flattening of all curves, the
effect being more pronounced for large mass ratios. A more systematic analysis
of simulations for different mass ratios will appear elsewhere
\cite{follow-ups}.

\section{Conclusions and outlook}

Our understanding of binary black hole mergers has recently been
revolutionized by the success of numerical relativity simulations. A partial
disappointment came from the remarkable simplicity of the
inspiral-merger-ringdown transition: the non-linearities of Einstein's theory
do not seem to leave spectacular imprints on the merger, that seems to be a
very short phase smoothly connecting the familiar ``inspiral'' and
``ringdown'' waveforms \cite{Buonanno:2006ui}.

In this paper we argued that non-linearities could show up in the {\em fine
  structure} of the ringdown waveform, as systematic time variations of the
ringdown frequencies and damping times. We also showed by explicit
calculations that such time variations {\em are} actually present in numerical
waveforms (see Figs.~\ref{performance}, \ref{performanceq} and \ref{wrQ}). The
variations we are looking for are typically $\sim 100$ times smaller than the
QNM frequencies themselves. This smallness calls for accurate parameter
estimation methods to extract ringdown frequencies from numerical simulations.
We considered a class of well-studied and robust parameter estimation methods
for complex exponentials in noise, which are modern variations of a linear
parameter estimation technique first introduced in 1795 by Prony. 

The comparison of different fitting methods can help resolve actual physical
effects from systematic parameter estimation errors, due to the variance and
bias of each particular fitting algorithm. For this reason we compared two
variants of the original Prony algorithm (the Kumaresan-Tufts and matrix
pencil methods \cite{kumaresantufts,matrixpencil}) against standard non-linear
least-squares techniques, such as the Levenberg-Marquardt algorithm. We found
that the two classes of methods have comparable variance, but Prony-type
methods tend to have slightly smaller bias. Prony methods have a number of
advantages with respect to standard non-linear least-squares techniques:

\begin{itemize}
\item[1)] They do not require an {\em initial guess} of the fitting
  parameters.

\item[2)] They provide us with a simple, efficient way to estimate QNM
  frequencies for the {\em overtones}, and even to estimate how many overtones
  are present in the signal \cite{reddybiradar,waxkailath}.

\item[3)] They are explicitly designed to deal with {\em complex signals}, so
  they should be most useful for ``generic'' waveforms, such as those produced
  by spinning, precessing black hole binaries. In the case of non-trivial
  polarization of the gravitational radiation it might become crucial to fit
  simultaneously the real and imaginary parts of the signal to exploit
  optimally the information carried by both degrees of freedom.
  
\item[4)] {\em Statistical properties} of Prony-based methods in the presence
  of noise (such as their variance and bias) are well studied and under
  control. Our Monte Carlo simulations suggest that the statistical properties
  of Prony-like methods as we vary the SNR are more consistent than for
  non-linear least-squares methods, even when we consider a single damped
  sinusoid (the dotted blue lines in Fig.~\ref{fig:var-bias} have a very
  irregular behavior).

\end{itemize}

The methods introduced in this paper should be useful both theoretically and
experimentally. From a theoretical standpoint, besides helping in the search
for non-linearities, Prony methods can also be used as ``diagnostic tools'' for
numerical simulations. For example, monitoring the time variation of ringdown
parameters from different multipolar components of the radiation we can check
that the end-product of a merger {\em is} indeed consistent with the Kerr
solution. An analysis of presently available numerical waveforms based on
Prony methods is ongoing \cite{follow-ups}.  

So far, Prony algorithms have generally been overlooked by the GW data
analysis community. We believe that the techniques presented in this paper
should prove useful to extract science from noisy ringdown signals in the
(hopefully not too distant) future, when GW detection will finally become a
reality.

\section*{Acknowledgements}
We are grateful to E.~H.~Djermoune, R.~Kumaresan, S.~L.~Marple, B.~Porat and
G.~Smyth for very useful correspondence and for sharing with us their
numerical routines. We also thank B.~Br\"ugmann, J.~Cardoso, M.~Hannam and
S.~Husa for useful discussions. VC acknowledges financial support from Funda\c
c\~ao Calouste Gulbenkian through the Programa Gulbenkian de Est\'{\i}mulo \`a
Investiga\c c\~ao Cient\'{\i}fica. Computations were performed at HLRS
(Stuttgart) and LRZ (Munich).  This work was supported in part by DFG grant
SFB/Transregio 7 "Gravitational Wave Astronomy", by Funda\c c\~ao para a
Ci\^encia e Tecnologia (FCT) -- Portugal through project PTDC/FIS/64175/2006,
by the National Science Foundation under grant number PHY 03-53180, and by
NASA under grant number NNG06GI60 to Washington University. J.G. and U.S.
acknowledge support from the ILIAS Sixth Framework Programme.


\end{document}